\documentstyle[12pt,amssym,aaspp4,psfig]{article}

\lefthead{Chiappini et al.}
\righthead{Oxygen, Carbon and Nitrogen evolution}

\begin{document}


\title{Oxygen, Carbon and Nitrogen evolution in galaxies}

\author{Cristina Chiappini\altaffilmark{1}}
\author{Donatella Romano\altaffilmark{2}}
\author{Francesca Matteucci\altaffilmark{3}}

\altaffiltext{1}{Osservatorio Astronomico di Trieste, 
		 Via G.B. Tiepolo 11, I-34131 Trieste, Italy;
		 chiappini@ts.astro.it}
\altaffiltext{2}{SISSA/ISAS, Via Beirut 2-4, I-34014 Trieste, Italy; 
       		 romano@sissa.it}
\altaffiltext{3}{Dipartimento di Astronomia, Universit\`a di Trieste,
                 Via G.B. Tiepolo 11, I-34131 Trieste, Italy; 
		 matteucci@ts.astro.it}





\begin{abstract}
We discuss the evolution of oxygen, carbon and nitrogen in galaxies
of different morphological type by adopting detailed chemical 
evolution models with different star formation histories 
(continuous star formation or starbursts). 
In all the models detailed nucleosynthesis prescriptions from 
supernovae of all types and low- and intermediate-mass stars
are taken into account.
We start by computing chemical evolution models for the
Milky Way with different stellar nucleosynthesis
prescriptions. Then, a comparison between model results and
``key'' observational constraints allows us to
choose the best set of stellar yields. 
Once the best set of yields is identified for the Milky Way, 
we apply the same nucleosynthesis prescriptions to 
other spirals (in particular M101) and dwarf irregular galaxies.
We compare our model predictions with the 
[C,N,O/Fe] vs. [Fe/H], log(C/O) vs. 12+ log(O/H),
log(N/O) vs. 12+ log(O/H) and [C/O] vs. [Fe/H] relations 
observed in the solar 
vicinity and draw the following conclusions: i) there is no need to invoke 
strong stellar winds in massive stars in order to explain the 
evolution of the C/O ratio, as often claimed in the literature; 
ii) the predicted [O/Fe] ratio as a function of metallicity is in very good
agreement with the most recent data (Mel\'endez \& Barbuy 2002) available
for the solar vicinity, especially for halo stars. This fact again suggests
that the oxygen stellar yields in massive stars computed either by Woosley \& 
Weaver (1995) or Thielemann et al. (1996) without taking into account
mass loss, well reproduce the observations;
iii) we predict that the gap observed in the [Fe/O] and 
[Fe/Mg] at [Fe/H]$\sim-$0.3 dex, should be observed also in 
C/O and N/O versus O/H. 
The existence of such a gap is predicted by our model for the Milky Way 
and is due to a halt in the star formation between the end of the thick disk 
and the beginning of the thin disk phase. Such a halt is produced by the 
adopted threshold gas density for the star formation rate; 
iv) this threshold is also
responsible for the prediction of a very slow chemical enrichment between 
the time of formation of the solar system (4.5 Gyr ago) and the present time,
in agreement with new abundance measurements; v)
the chemical evolution models for dwarf irregulars and spirals, 
adopting the same nucleosynthesis prescriptions of the best model for the
solar neighbourhood, well reproduce the available constraints for these objects; vi) by taking into account the results obtained for all the studied galaxies (Milky Way, M101, dwarf galaxies and DLAs) we 
conclude that there is no need for claiming a strong primary component of N 
produced in massive stars (M$>$10M$_{\odot}$);
vii) moreover,
there is a strong indication that C and N are mainly produced in 
low- and intermediate-mass stars, at variance with recent suggestions
that most of the C should come from massive stars.
In particular, intermediate-mass stars
with masses between 4 and 8 M$_{\odot}$ contribute mostly to N (both
primary and secondary) whereas 
those with masses between 1 and 3 M$_{\odot}$ contribute mostly to C. 
At the same time, our results suggest that
the C yields computed for massive stars, without taking 
into account stellar rotation, are underestimated and 
should be at least a factor of 3 larger than the current values.

\end{abstract}


\keywords{Galaxy: formation --- Galaxy: evolution --- Galaxy: 
abundances, nuclear reactions, nucleosynthesis, abundances -- galaxies: individual: 
M101 -- galaxies: irregular}


%

\section{Introduction}

The variation of element abundance ratios with metallicity
can be used as a powerful tool for understanding the
chemical enrichment of galaxies and hence, their evolution. 
Given the fact that different chemical elements are produced on different
timescales by stars of different lifetimes,
the abundance ratios of some key elements versus metallicity
will depend not only on the stellar evolution processes 
but also on the star formation history
(SFH) of a galaxy (e.g., Pagel 1997). The SFH in turn depends on 
important processes taking place during the 
galaxy evolution like, for instance, inflow and outflow of gas.

Because of
their long main-sequence lifetimes and lack of 
deep convective zones, the lower mass stars which are still observable
today, have preserved the patterns
of elemental abundances generated by the initial
burst of star formation. In principle, by analyzing
the chemical abundances of stars of various ages,
lying at different positions inside a galaxy, it is possible
to constrain its star formation and enrichment histories.
However, since high quality spectra
of individual stars are required to determine abundances,
studies of this kind are possible only in 
a few cases.

The CNO elements, and in particular the C/O and N/O abundance ratios, 
can be considered ``key'' tools for the study of chemical
evolution of galaxies, as these are elements produced by different mechanisms
and in different stellar mass ranges. Oxygen is almost entirely produced by 
massive stars and ejected into the interstellar medium via the explosion
of type II SNe. The situation for C and N is more complex
as these elements can be produced by stars of all masses.
N is mostly a secondary element being a product of the 
CNO cycle and formed at expenses of the C and O already present
in the star, although a primary N component originating in asymptotic giant
branch (AGB) stars
is also predicted by stellar nucleosynthesis studies.
In fact, primary nitrogen can be produced during the third
dredge-up, occurring along the AGB phase, if nuclear burning
at the base of the convective envelope is efficient (hot-bottom
burning, HBB, Renzini \& Voli 1981 - hereinafter RV). 
Some primary N can also be produced in massive stars due to 
stellar rotation, according to the recent calculations of 
Meynet \& Maeder (2002b). Carbon is a primary element 
produced during the quiescent He-burning in massive and 
intermediate-mass stars. However, there are 
many uncertainties still involved in the carbon yields.

The abundance ratio of a secondary to a primary element is predicted
to increase with the abundance of its seed. 
This is the case for the N/O ratio as N is mostly a secondary element,
especially in massive stars. On the other hand,
the abundance ratio of two primary elements such as $^{12}$C and $^{16}$O, 
which are restored into the ISM by stars
in different mass ranges, shows almost the same behavior with 
metallicity as the ratio between a secondary and a primary
element: the C/O ratio in fact increases with metallicity.
The most straightforward interpretation is that 
$^{12}$C is mainly restored into the ISM by 
intermediate-mass stars (and hence on longer timescales compared
to the $^{16}$O enrichment which comes mainly
from massive stars), so that the C/O ratio
increases as the ISM enrichment proceeds. 
In other words, the production of a primary element 
on long timescales mimics the secondary nature.
This interpretation
has been recently challenged by several papers (eg. Henry et al. 2000,
Carigi 2000 - see below). 

The best place for studying the evolution of the CNO elements
is the Milky Way (MW), where it is possible to 
measure the abundances of stars of different
ages and hence infer the chemical composition of the 
ISM at various epochs. In particular, the abundances
of the CNO elements in metal-poor stars provide an important
constraint on the early chemical evolution of the Milky Way.
The abundances of the very-metal-poor stars belonging to the halo
trace the composition
of the ISM at the time of their formation, i.e., many billion years ago,
when most of the enrichment was due to massive stars.
B stars and HII regions trace instead the ISM chemical composition
in the Galactic disk at the present time. 
In the past years a great deal of observational work has been
devoted to measure stellar abundances in the Milky Way, both
in halo and disk stars (for a large compilation of the available
abundance data until 1999 see Chiappini et al. 1999).
More recently, careful and detailed
abundance analyses have been carried out by several groups
(e.g., Carretta et al. 2000; Fuhrmann 1998; Mel\'endez et al. 2001,
Mel\'endez \& Barbuy 2002, Depagne et al. 2002, Nissen et al. 2002). 
In particular, a few
high quality data are now available for C and O at low metallicities  
(Carretta et al. 2000). 
 
For the Milky Way, the data reveal that oxygen
(as well as other $\alpha$-elements) shows an 
overabundance relative to Fe in metal-poor 
stars ([Fe/H]$<$ $-$1.0), whereas these ratios decrease 
for disk stars until they reach the solar value.
This trend is generally 
interpreted as due to the time-delay in the iron production which 
originates from long living white dwarfs in binary systems 
eventually exploding as type Ia SNe.
The available data for Mg, Si, Ca and S show that for [Fe/H] $<$ $-$1.0 the
[$\alpha$/Fe] ratio is almost constant, defining a plateau, 
whereas for oxygen 
there seems to be a slight increase of the [O/Fe] ratio with 
decreasing metallicity (see Chiappini et al. 2001 for a discussion 
on this particular point - hereafter CMR2001).
On the other hand, recent papers (Israelian et al. 1998, 2001;
Boesgaard et al. 1999), show an even steeper increase of the [O/Fe] ratio 
with decreasing [Fe/H]. This trend, if real, is difficult to reconcile 
with our knowledge about stellar nucleosynthesis. In the present work we 
discuss carefully the abundance data concerning oxygen and show that
when only the most reliable measurements and abundance analysis (Asplund
\& Garcia P\'erez 2001 - see Sect. 4.1.2) are considered, a good agreement 
with our theoretical predictions is found.

As far as C and N are concerned, there are still many open questions. 
It has been shown that chemical evolution models
for the MW, adopting the yields computed by RV
for low- and intermediate-mass stars together with
yields for massive stars as computed by Woosley and
collaborators (Woosley et al. 1984; Woosley \& Weaver 1995), 
could not reproduce the steep rise of C/O vs. O/H 
observed in the solar vicinity (Garnett et al. 1999). 
A way to solve this problem was suggested by Prantzos et al. (1994)
who adopted the yields computed by Maeder (1992) which assume
strong mass loss by stellar winds in massive stars. The C yields predicted by 
Maeder (1992) increase with metallicity, because mass loss itself is an
increasing function of metallicity. Prantzos et al. showed that this
C/O increase as well as the solar value of this ratio, could be reproduced
by models adopting RV yields for low- and 
intermediate-mass stars and Maeder (1992) ones for massive stars. 
This fact, together with the uncertainty related to the 
$^{12}$C($\alpha$,$\gamma$)$^{16}$O reaction,
led to the view that the main contributors
to the carbon we observe today in the ISM are massive stars.
Recent work in the literature seems to confirm this 
suggestion, even when the more recent yields of van den Hoek \& Groenewegen
(1997 - hereafter vdHG) for low- and intermediate-mass stars are used instead
of those of RV (Liang et al. 2001; Henry et al. 2000; Carigi 2000).
In the present paper we present strong arguments against such
interpretation and we suggest that both the N and the C  
we observe at the present time
in the ISM were mostly produced inside low- and intermediate-mass stars.
Moreover we show that the N data in the solar vicinity can be explained
without invoking important quantities of
primary N in massive stars.
We then check if 
our conclusions on the evolution of C, N and O in the
MW are consistent with the CNO data available 
for other galaxies (blue compact galaxies - BCGs, other spiral galaxies and 
Damped Lyman-$\alpha$ systems - DLAs), under the assumption that 
the stellar nucleosynthesis should be the same for all galaxies. 

Abundance data are available for some extragalactic HII regions 
(Garnett et al. 1995a,b; 1997a,b; van Zee et al. 1997), 
including those measured in the outer 
parts of spiral disks (Ferguson et al. 1998; van 
Zee et al. 1998a,b; Garnett et al. 1999) and
BCGs (Izotov et al. 1999; Izotov \& Thuan 1999). 
Some of these objects have metallicities down to 
around $\simeq$ 1/10 solar and are useful to study
the behavior of chemical abundances at low metallicities.
This is especially true for elements such as C and N. 

The N/O vs. O/H diagram for dwarf galaxies
is often interpreted in the literature (the same is true
for the C/O vs. O/H diagram) as an evolutionary diagram, but instead
it represents the final abundance values achieved by
objects which evolved in a completely different way from each other (Diaz \&
Tosi 1986, Matteucci \& Tosi 1985).
Therefore, a meaningful comparison between model predictions
relevant to dwarf irregulars and data should involve only the
end points of the theoretical evolutionary tracks.
Only in the case of the stars in the Milky Way we face an evolutionary
diagram, where O/H can be interpreted as a time-axis.

One of the main questions nowadays is to assess the nature 
of nitrogen production in massive stars
in order to explain the observations of N/O ratios in dwarf galaxies
and DLAs which have low metallicities (Pilyugin 1999). The small ``plateau'' 
in the N/O ratio 
observed in low metallicity HII regions (e.g. Izotov and Thuan 1999) 
is often quoted as one indication
that massive stars should produce an important quantity of primary N. 
This seems to be in conflict with the 
fact that the nitrogen over Si or S in some DLAs are well
below the typical value observed in low-metallicity BCGs
(e.g., Lu et al. 1998). In fact, DLAs offer another important
piece of information (Pettini et al. 1995; Prochaska \& Wolfe 2002)
as in this case it is possible to probe the {\it very early phases} 
of evolution of such systems. In this paper we show that the dwarf 
galaxy and DLA data can be understood without the necessity for 
primary nitrogen from massive stars.

Clearly, the best way to determine the lower limit of the N/O ratio
and hence the existence of a possible primary N contribution
from massive stars would be to measure the N/O ratios
in Galactic halo stars at metallicities below [Fe/H] $\simeq-$2.
As we will show, this corresponds to the first 30 Myr of the 
evolution of the Milky Way, when the ISM was still not 
enriched by any intermediate-mass stars (the lifetime
of an 8 M$_{\odot}$ star is around 30 Myr). Unfortunately,
the available data for N in halo stars are still 
too uncertain. 

Finally, we discuss the problem of the abundance pattern
in outer spiral disks. The CNO abundance ratios in outer
disks are similar to those observed in dwarf irregular
galaxies (van Zee et al. 1998b). We argue that this 
is consistent with the concept of ``inside-out'' formation of 
the disk, in which the timescale for accretion of gas 
onto the forming disk increases radially outwards 
(Matteucci \& Fran\c cois 1989; Chiappini et al. 1997, 
hereafter CMG97; CRM2001).
In such a scenario, outer spiral disks have experienced
slow star formation like dwarf irregulars.
However, some important differences exist and we will address
them by studying one spiral galaxy in particular, M101.

The paper is organized as follows. In Section 2 we discuss the different yield
sets which we then use in the chemical evolution model for
the MW. In Section 3 we present our chemical evolution models for 
the MW, dwarf galaxies and for M101. In Section 4 our results
are shown and the conclusions are drawn in Section 5.

\section{Stellar yields}

In this section we describe the stellar yields adopted in our chemical
evolution models. In particular, for
low- and intermediate-mass stars we adopted the yields of RV for two
different efficiencies of HBB,
and those of vdHG for two different cases of mass loss
during the AGB phase (see below). For massive stars we adopted both the yields
of Woosley \& Weaver (1995 - hereafter WW) and Nomoto et al. (1997) 
(these latter are essentially the same as in Thielemann, Nomoto 
\& Hashimoto 1996 - heareafter TNH, but for an enlarged grid of masses). 
None of these models 
accounts for mass loss by stellar winds and/or stellar rotation, although 
these processes could affect the chemical yields.

When concluding this paper, new stellar calculations appeared.  
Those are the new computations of 
Meynet and Maeder (2002b) and
will be included in a future paper. However, the main differences with respect to the 
yields adopted here will be discussed in the next sections. 

\subsection{Low- and intermediate-mass stars}

Stars with initial masses between $\sim$\,1 and $\sim$\,5\,--\,8$M_\odot$ (depending on 
stellar evolution models) experience a phase of double-shell burning at the end 
of their life, referred to as AGB phase. During this 
phase, they eject into the ISM significant amounts of 
$^4$He, $^{12}$C, $^{13}$C, and $^{14}$N.
The ejected masses reflect important abundance variations occurred 
when the stars were on the AGB phase.

In vdHG the stellar yields were computed by 
taking into account several new physical ingredients (see Groenewegen \& de 
Jong 1993; Groenewegen et al. 1995) and the detailed 
treatment of mass loss and chemical evolution prior to the AGB. Moreover, the 
algorithms used are metallicity dependent. This is an important point, since 
mass-loss rates vary strongly with 
initial metallicity (e.g., Schaller et al. 1992); moreover, observations show 
that the luminosity function and the relative number ratios of carbon- and 
oxygen-rich AGB stars in galaxies of different metallicities are different 
(e.g., Groenewegen \& de Jong 1993).

In the vdHG paper, the theoretical yields were computed for stars with 
initial masses between 
$\sim$\,0.8 and $\sim$\,8 $M_\odot$ and initial metallicities $Z$ = 0.001, 
0.004, 0.008, 0.02, and 0.04. Those authors followed in detail the stellar 
evolution and mass loss up to the end of the AGB. They included in their 
computations the first, second, and third dredge-up phases and the effect of 
HBB, though in an approximate way. The free parameters 
(the mass loss scaling parameter $\eta_{AGB}$ for stars on the AGB, the 
minimum core mass for dredge-up, and the third dredge-up efficiency) were 
fixed to obtain the best agreement with observations of AGB 
stars in both the Galaxy and the Large Magellanic Cloud (LMC). In particular, 
for the mass loss scaling parameter, a value of $\eta_{AGB}$ = 4 was adopted. 
However, for AGB stars in low metallicity systems, values of $\eta_{AGB}$ 
$\sim$\,1\,--\,2 may be more appropriate (arguments are given in Groenewegen 
et al. 1995). With decreasing values of $\eta_{AGB}$ (i.e., smaller mass-loss 
rates), the resulting yields increase owing to the longer AGB lifetimes which 
favor more thermal pulses. We computed chemical evolution models for both cases: 
{\it i)} $\eta_{AGB}=$4 constant with metallicity (adopted in the models of 
Carigi 2000 and Henry et al. 2000), and {\it ii)} $\eta_{AGB}$ 
varying from 1 to 4 with increasing metallicity.

During the third dredge-up, carbon is dredged-up to the stellar surface and 
the star eventually becomes a C-star. For stars with initial masses 
larger than 3$\,-\,$4 $M_\odot$, 
this transition is affected by HBB. Both the carbon already present and the newly 
dredged-up one are processed at the base of the convective envelope through 
the CNO cycle. Unfortunately, details of this process are not well understood. 
RV treated HBB in considerable detail as a function of the mixing length 
parameter $\alpha$. We give results of chemical evolution models adopting 
their yields in the case $\alpha$ = 0 (no HBB) and $\alpha$ = 1.5. In the standard 
model of vdHG, HBB is included at a level consistent with the $\alpha$ = 2 
case of RV, i.e., the largest HBB effect, which according to the authors, agrees 
with both 
theory (Boothroyd et al. 1993, 1995) and observations (Plez et al. 1993; Smith 
et al. 1995). Under these conditions, HBB operates in stars with $M > 4 
M_\odot$. 

Summarizing the main nucleosynthesis results concerning low- and 
intermediate-mass stars:
the carbon yields increase with decreasing initial metallicity, owing to the 
fact that dredge-up and subsequent CNO-burning affect more strongly the 
composition of envelopes with low metallicity. 
Moreover, the core mass 
at the first thermal pulse is larger at low metallicity, thus resulting in a 
larger amount of material dredged-up to the envelope. 
On the contrary, the nitrogen yields slightly increase with metallicity, 
as nitrogen is formed by consumption of C and O already present
in the stars (secondary production) and by HBB during the AGB phase (primary production). 
Oxygen production from low- and intermediate-mass stars is practically irrelevant. 
Note that a non-negligible quantity of C is also produced by Type Ia SNe and this
was included in our calculations (for Type Ia SNe we adopted the yields of 
Thielemann et al. 1983).

\subsection{Massive stars}

The oxygen and the other $\alpha$-elements (Ne, Mg, Si, S, Ca and Ti)
are mainly produced during the hydrostatic burning phases in massive stars
($M \ge 10 M_{\odot}$). Their yields depend crucially
on the pre-supernova model (convection criterion, mixing processes, 
mass loss and nuclear reaction rates).
The oxygen and the other $\alpha$-elements are normally restored into the 
ISM via SN explosion (type II SNe). Stellar winds, 
even in Wolf-Rayet stars, are not able to carry oxygen away but the effect 
of mass loss is to increase the amount of lost helium and therefore to 
decrease the final 
amount of oxygen (since He is later transformed into heavy elements). 
Thus the net effect of a large mass loss is to 
increase the yields of He and decrease those of oxygen (see Maeder 1992).
Oxygen is not affected by explosive nucleosynthesis, whereas Si, S and Ca are:
these elements are also produced in non-negligible amounts
during the explosion of SNe Ia (the outcome of
C-deflagration in C-O white dwarfs). On the other hand, O and Mg are 
almost entirely produced in massive stars.
The Fe-peak elements are mostly formed during explosive nucleosynthesis, 
mainly in type Ia SNe.
In the explosive nucleosynthesis, the yields depend on the uncertain speed 
of the burning front and convection and, for type II SN explosions, on the
``mass-cut''.

Here we recall the average O and Fe yields from massive stars, obtained 
by integrating on the Salpeter initial mass function (IMF), 
as derived from the most recent calculations 
(WW; TNH), as well as the yields of O and Fe from 
type Ia SNe (Thielemann et al. 1993).
These yields are:
$<y_{O}>_{SNII}= 1.77-0.59 M_{\odot}$,
$<y_{O}>_{SNIa}= 0.143 M_{\odot}$,
$<y_{Fe}>_{SNII}= 0.14-0.07 M_{\odot}$, $<y_{Fe}>_{SNIa}= 0.744 M_{\odot}$.
They clearly show that the bulk of O should originate from type II SNe, 
whereas the bulk of Fe should originate from type Ia SNe when a Salpeter
like IMF is chosen.

The effects of $Z$-dependent yields including mass loss on Galactic chemical 
evolution have been studied by, e.g., Maeder (1992) and Prantzos et al. (1994). 
Prantzos et al. (1994) concluded that the growth of 
the C yields from massive stars with increasing metallicity is a key point in 
order to properly understand the [C/O] vs. [Fe/H] relation of disk stars. 
However, a recent study (Meynet \& Maeder 2002b)
seems to point to a less important effect of mass-loss than the one
given in Maeder (1992). On the other hand, these new results suggest
that rotation can increase the C yields, especially at Z $>$ 0.004. 
However, Meynet \& Maeder (2002b) do not include a self-consistent calculation
of the Fe yields, therefore we do not use these yields. 
In fact, to be able to test these new calculations on chemical evolution 
models, a complete grid of stellar yields for several elements is needed.
However, in order to take in account, although in an approximate way, 
the rotation and mass loss effects we  
also run some models in which the C yields from WW and TNH are multiplied by a 
factor of 3 in the mass range 40-100 $M_\odot$ (see Sect. 2.3).

\subsection{Comparison of the different yield sets}

In Table 1 we present the yields per stellar generation (i.e., integrated 
over the IMF - see Henry et al. 2000). 
These quantities are listed for two different IMF choices. 
For $^{14}$N we show the total yield (primary+secondary). From this table 
we can see for instance: 
{\it i)} the effect of the $\alpha$ parameter on the stellar yields computed by RV,
where models with $\alpha$=0 (no HBB) predict less N and more C (as well
as a little more O, although the contribution to O enrichment by IMS is very low);
{\it ii)} the effect of a high $\eta_{AGB}$ parameter on the yields of vdHG
(case $\eta_{AGB}$=4 for all metallicities), in a situation of maximum
efficiency for HBB. In this case, the C yields dependence on metallicity is small;
{\it iii)} when adopting $\eta_{AGB}$ smaller at lower metallicities (case 
$\eta_{AGB}$ var) the C yields are much larger for Z=0.001 and Z=0.004 
and decrease with metallicity;
{\it iv)} for massive stars the secondary N production predicted by WW
is larger than the one of TNH, while for O the opposite happens.
For C both results are similar.

In Figure 1 we show a comparison of different carbon yields for massive stars, computed
for solar metallicity (open circles: TNH; 
filled circles: TNH with carbon yields multiplied by 3;
open triangles: WW; filled triangles: WW with carbon yields
multiplied by 3; stars: Meynet \& Maeder 2002b).
The first thing to notice in this figure is that WW and TNH yields
multiplied by a factor of 3 in the 40-100 M$_{\odot}$ range
are similar to the new calculations of Meynet \& Maeder (2002b).
In the lower mass range all the yield sets give quite similar results
(and in fact we did not multiply the carbon yields for M $<$ 40 M$_{\odot}$).
However, an important difference exists in the 9-13 M$_{\odot}$ mass range. In fact,
the Meynet \& Maeder yields are larger than the ones given both in TNH and WW.
This would certainly lead to a better agreement with the observed C/O solar value,
than the one obtained in the present paper. 

\begin{deluxetable}{ccccccc}
\small
\tablecolumns{7}
\renewcommand{\tabcolsep}{4pt}
\renewcommand{\arraystretch}{.6}
\tablewidth{0pt}
\tablenum{1}
\tablecaption{\small{Integrated yields.}}
\tablehead{
\colhead{$Z$} & {P$_{^{12}C}$} & {P$_{^{14}N}$} & {P$_{^{16}O}$} & {mass range}
& {IMF slope} & {ref.}
}
\startdata
0.001 &	6.7E-4 & 5.3E-4 & 7.8E-5 & 0.9\,--\,8.0 & Salpeter & vdHG 
($\eta_{AGB}$ cost) \nl
0.004 &	7.3E-4 & 5.7E-4 & 6.0E-5 & 0.9\,--\,8.0 & Salpeter & vdHG 
($\eta_{AGB}$ cost) \nl
0.008 &	6.1E-4 & 6.2E-4 & 2.4E-5 & 0.9\,--\,8.0 & Salpeter & vdHG 
($\eta_{AGB}$ cost) \nl
 0.02 &	4.1E-4 & 7.2E-4 & 1.2E-5 & 0.9\,--\,8.0 & Salpeter & vdHG 
($\eta_{AGB}$ cost) \nl
 0.04 &	1.2E-4 & 9.1E-4 & 1.1E-5 & 0.9\,--\,8.0 & Salpeter & vdHG 
($\eta_{AGB}$ cost) \nl
0.001 &	1.9E-3 & 9.6E-4 & 2.2E-4 & 0.9\,--\,7.0 & Salpeter & vdHG 
($\eta_{AGB}$ var) \nl
0.004 &	1.3E-3 & 7.4E-4 & 1.2E-4 & 0.9\,--\,7.0 & Salpeter & vdHG 
($\eta_{AGB}$ var) \nl
0.008 &	6.1E-4 & 6.2E-4 & 2.4E-5 & 0.9\,--\,8.0 & Salpeter & vdHG 
($\eta_{AGB}$ var) \nl
 0.02 &	4.1E-4 & 7.2E-4 & 1.2E-5 & 0.9\,--\,8.0 & Salpeter & vdHG 
($\eta_{AGB}$ var) \nl
 0.04 &	1.2E-4 & 9.1E-4 & 1.1E-5 & 0.9\,--\,8.0 & Salpeter & vdHG 
($\eta_{AGB}$ var) \nl
0.001 &	6.6E-4 & 4.0E-4 & 7.3E-5 & 0.9\,--\,8.0 & Scalo & vdHG ($\eta_{AGB}$ 
cost) \nl
0.004 &	7.1E-4 & 4.3E-4 & 5.9E-5 & 0.9\,--\,8.0 & Scalo & vdHG ($\eta_{AGB}$ 
cost) \nl
0.008 &	6.1E-4 & 4.8E-4 & 2.9E-5 & 0.9\,--\,8.0 & Scalo & vdHG ($\eta_{AGB}$ 
cost) \nl
 0.02 &	3.9E-4 & 5.9E-4 & 2.8E-5 & 0.9\,--\,8.0 & Scalo & vdHG ($\eta_{AGB}$ 
cost) \nl
 0.04 &	6.5E-5 & 7.9E-4 & 3.1E-5 & 0.9\,--\,8.0 & Scalo & vdHG ($\eta_{AGB}$ 
cost) \nl
0.001 &	1.9E-3 & 7.6E-4 & 2.1E-4 & 0.9\,--\,7.0 & Scalo & vdHG ($\eta_{AGB}$ 
var) \nl
0.004 &	1.3E-3 & 5.8E-4 & 1.2E-4 & 0.9\,--\,7.0 & Scalo & vdHG ($\eta_{AGB}$ 
var) \nl
0.008 &	6.1E-4 & 4.8E-4 & 2.9E-5 & 0.9\,--\,8.0 & Scalo & vdHG ($\eta_{AGB}$ 
var) \nl
 0.02 &	3.9E-4 & 5.9E-4 & 2.8E-5 & 0.9\,--\,8.0 & Scalo & vdHG ($\eta_{AGB}$ 
var) \nl
 0.04 &	6.5E-5 & 7.9E-4 & 3.1E-5 & 0.9\,--\,8.0 & Scalo & vdHG ($\eta_{AGB}$ 
var) \nl
 0.02 & 6.6E-4 & 1.0E-3 & -5.5E-5 & 1.0\,--\,8.0 & Salpeter & RV ($\alpha$ = 
1.5) \nl
 0.02 & 5.6E-4 & 7.6E-4 & -4.0E-5 & 1.0\,--\,8.0 & Scalo & RV ($\alpha$ = 1.5) 
\nl
 0.02 & 1.3E-3 & 1.8E-4 & -1.8E-5 & 1.0\,--\,8.0 & Salpeter & RV ($\alpha$ = 
0) \nl
 0.02 & 1.1E-3 & 1.6E-4 & -1.4E-5 & 1.0\,--\,8.0 & Scalo & RV ($\alpha$ = 0) 
\nl
 0.02 & 4.9E-4 & 8.8E-6 & 1.3E-2 & 13.0\,--\,70.0 & Salpeter & TNH \nl
 0.02 & 2.1E-4 & 4.2E-6 & 4.9E-3 & 13.0\,--\,70.0 & Scalo & TNH \nl
 0.02 & 5.9E-4 & 1.8E-4 & 6.2E-3 & 11.0\,--\,40.0 & Salpeter & WW \nl
 0.02 & 2.8E-4 & 8.6E-5 & 2.7E-3 & 11.0\,--\,40.0 & Scalo & WW \nl
\enddata
\end{deluxetable}

In Tables 2a and 2b we list the main nucleosynthesis prescriptions used in our 
models for the Milky Way. The second column indicates the adopted yields 
for low- and intermediate-mass
stars, the third column indicates those for massive stars. The last column
indicates those models where the C yields were increased by a factor of 3 in stars 
in the 40 to 100M$_{\odot}$ mass range. 
Table 2b shows models analogous to model 6 but with modified nucleosynthesis for what
concerns the prescriptions for N (model 8) and without a gas density threshold in the disk (model 9 - Sect. 3.1). In particular, for model 8 we assumed an important primary N contribution
from massive stars.

\begin{deluxetable}{cccc}
\small
\tablecolumns{4}
\renewcommand{\tabcolsep}{4pt}
\renewcommand{\arraystretch}{.6}
\tablewidth{0pt}
\tablenum{2a}
\tablecaption{\small{Nucleosynthesis prescriptions.}}
\tablehead{
\colhead{Model} & \colhead{low- and intermediate-mass stars} & 
\colhead{massive stars} & \colhead{3 $\times$ $^{12}C$ in the range 
40\,--\,100 $M_\odot$}
}
\startdata
1  & RV ($\alpha$ = 0)        & TNH &  no \nl
2  & RV ($\alpha$ = 1.5)      & TNH &  no \nl
3  & vdHG ($\eta_{AGB}$ cost) & TNH &  no \nl
3a & vdHG ($\eta_{AGB}$ cost) & TNH & yes \nl
4  & vdHG ($\eta_{AGB}$ var)  & TNH &  no \nl
5  & vdHG ($\eta_{AGB}$ var)  & WW  &  no \nl
6  & vdHG ($\eta_{AGB}$ var)  & TNH & yes \nl
7  & vdHG ($\eta_{AGB}$ var)  & WW  & yes \nl
\enddata
\end{deluxetable}

\begin{deluxetable}{cccc}
\small
\tablecolumns{4}
\renewcommand{\tabcolsep}{4pt}
\renewcommand{\arraystretch}{.6}
\tablewidth{0pt}
\tablenum{2b}
\tablecaption{\small{Nucleosynthesis prescriptions.}}
\tablehead{
\colhead{Model} & \colhead{Low- and intermediate} & \colhead{Massive stars} & \colhead{Modification}\\
 & \colhead{mass stars} & { } & { } }
\startdata
8  & vdHG ($\eta_{AGB}$ var)  & TNH & +primary N for massive stars \nl
9 & vdHG ($\eta_{AGB}$ var)  & TNH & no threshold in the disk \nl
\enddata
\end{deluxetable}

\section{The chemical evolution models}

\subsection{The chemical evolution model for the Milky Way}

We adopt the two-infall model of CMG97 (see also CMR2001) where the formation
of the halo is almost disentangled from that of the disk.
The halo and bulge form on a relatively short timescale (0.8-1.0 Gyr)
out of a first infall episode, whereas the disk accumulates
much more slowly and ``inside-out'' during a second independent
infall episode.
The Galactic disk is approximated by 
several independent rings, 2 kpc wide, without exchange of matter between 
them. The rate of accretion of matter in each shell is:
\begin{equation}
\frac{d\Sigma_I(R, t)}{dt} = A(R)\,e^{- t/\tau_{H}} + B(R)\,e^{- (t - 
t_{max})/\tau_{D}},
\end{equation}
where $\Sigma_I(R, t)$ is the surface mass density of the infalling material, 
which is assumed to have primordial chemical composition; $t_{max}$ is the 
time of maximum gas accretion onto the disk, coincident with the end of the 
halo/thick-disk phase and set here equal to 1 Gyr; $\tau_{H}$ and 
$\tau_{D}$ are the timescales for mass accretion onto the halo/thick-disk and 
thin-disk components, respectively. In particular, $\tau_{H}$ = 0.8 Gyr and, 
according to the ``inside-out'' scenario, $\tau_{D}(R)$ = 1.033$\times 
(R$/kpc)$-$1.267 Gyr, in order to have $\tau_{D}($8kpc$)$=7 Gyr as required
to fit the G-dwarf metallicity distribution. We adopt a linear approximation 
for the variation of $\tau_{D}(R)$. This variation is constrained by 
the observed radial profiles of gas, star formation rate (SFR) and abundances. In some cases the 
observed radial profiles suggest that this timescale can approach a constant 
value for Galactocentric distances larger than $\simeq$ 12 kpc. 
The quantities $A(R)$ and $B(R)$ are derived from the condition of 
reproducing the current total surface mass density distribution in the halo 
and along the disk, respectively (Rana 1991). 

The SFR adopted here has the same formulation as in CMG97: 

\begin{equation}
\psi(R, t) = \nu(t)\,\left( \frac{\Sigma(R, t)}{\Sigma(R_\odot, t)} 
\right) ^{2\,(k - 1)}\,\left( \frac{\Sigma(R, t_{Gal})}{\Sigma(R, t)} \right)
^{k - 1}\,G^{k}(R, t),
\end{equation}
where $\nu(t)$ is the efficiency of the star formation process, $\Sigma(R,t)$ 
is the total surface mass density at a given radius $R$ and a given time $t$, 
$\Sigma(R_\odot, t)$ is the total mass surface density at the solar position, 
$\Sigma_{gas}(R, t)$ is the gas surface mass density and $G$ is the normalized
gas density, i.e., $\Sigma_{gas}(R, t)/\Sigma(R, t_{Gal})$. Note that the gas surface 
density exponent, $k=$1.5, was obtained from the best model of CMG97 
in order to ensure a good fit to the observational constraints at the solar 
vicinity. This value then turned out to be in very good agreement with 
observational results by Kennicutt (1998) and with N-body simulation results 
by Gerritsen \& Icke (1997).

The efficiency of star formation is set to be $\nu$=1 Gyr$^{-1}$ to ensure the 
best fit to the observational features in the solar vicinity, and becomes zero 
when the gas surface density drops below a certain critical gas threshold 
(Kennicutt 2001). We adopt a threshold density 
$\Sigma_{th}$ $\sim$ 7 $M_\odot$ pc$^{-2}$ 
in the disk (CMG97). As far as the halo/thick-disk phase is concerned, a 
similar value for the threshold is expected (see Elmegreen 1999). 
The IMF is that of Scalo (1986), assumed to stay constant during the evolution 
of the Galaxy (see Chiappini, Matteucci \& Padoan 2000).
In particular, here we adopt Model A of CMR2001 which assumes
that the total halo/thick-disk mass density profile is constant
for $R$ $\le$ 8 kpc 
and decreases as $R^{-1}$ outwards. 

This model provides a very good fit to the G- and K-dwarf 
metallicity distributions in the solar vicinity (Kotoneva et al. 2002)
and requires a  timescale  
for the formation of the disk in the solar neighbourhood of $\simeq$7 Gyr.
The model also provides a good fit to 
the present time star formation rate, surface gas density, SN rates
as well as abundance gradients, gas and stellar distributions 
along the disk (see CMR2001).

\subsection{The chemical evolution model for dwarf irregular galaxies} 

The evolution of dwarf irregular galaxies is often assumed to be 
characterized by
a bursting star formation history. These low-mass
systems are particularly sensitive to outflows resulting
from the energy injection from the SN explosions. Moreover, they
could also suffer infall from the extended neutral
gas envelopes often observed around such galaxies
(see for example Matteucci and Chiosi 1983). 

In this work we adopt a model based on the work of Bradamante
et al. (1998) and take into account the recent chemodynamical 
results by Recchi et al. (2001).
In particular, we adopt the best prescription for the energetics
given in Recchi (2001) and Recchi et al. (2001): 
we assume a low thermalization efficiency
for type II SNe ($\eta_{II}=0.03$) and a maximun value for SNeIa
($\eta_{Ia}=1.0$); for the stellar winds we adopt a thermalization
efficiency of $\eta_{w}=0.03$. Infall is also considered
with a timescale $\tau$=5 $\times$ 10$^8$ yr and primordial chemical composition.
An important result found by
Recchi et al. (2001), with the above prescriptions, is 
that the ejecta of type Ia SNe
and intermediate-mass stars are lost from the parent
galaxy more easily than type II SN ejecta, so that the
Fe and N ejection efficiencies are larger than the ejection
efficiencies of the $\alpha$-elements (e.g. O, Mg).

The assumed model parameters are: a total luminous mass
at the present time of 10${^9}M_{\odot}$, a dark
to luminous matter ratio of 10 (which is rather a standard
value for BCGs as well as for IZw18 - see Recchi 2001).
We adopt the same nucleosynthesis prescriptions
as in the best model for the Milky Way (model 6, see next section). 
Moreover for these dwarf starbursting galaxies we adopt a slightly flatter IMF, 
as often suggested in the literature, especially for the more
metal-poor galaxies such as IZw18 (Aloisi et al. 1999), 
with an exponent of x=$-$1.1 (for comparison, the Salpeter 
IMF by mass has x=$-$1.35).

\subsection{The chemical evolution model for M101}

CNO abundances have been measured in a significant sample of spirals 
(Thurston et al. 1996; Garnett et al. 1997a; Ferguson et al. 1998; 
van Zee et al. 1998a,b; Garnett et al. 1999). 
This allows us to test the nucleosynthesis yields which 
successfully explain the CNO observations in stars of our own Galaxy against 
observations in others late-type galaxies.

We construct a chemical evolution model suited to follow the evolutionary 
history of a spiral galaxy more massive than the Milky Way, taking M\,101 as 
a template. M\,101 is assumed to form out of two main infall episodes, 
similarly to what happens for the Milky Way. The first infall episode forms 
the bulge, inner halo and thick disk substructures on a very short time-scale 
($\tau_{H}^{M\,101}$ = 0.7 Gyr). During the second infall episode the thin 
disk is formed, mainly through accretion of external matter of primordial 
chemical composition. The disk forms in an ``inside-out'' way, as in the
case of the MW. The adopted timescales are
$\tau_{D}^{M\,101}(R)$ = 1, 2.5, 5.5, 13 Gyr at a radius 
$R$ = 2, 4, 8, 18 kpc, respectively (see detailed discussion in Section 4.2). 
These timescales have been
chosen in order to reproduce the abundance gradients and the gas distribution
along the disk of M101.
The efficiency of star formation, $\nu$, the exponent of the 
star formation law, $k$, and the threshold in the gas density below which the 
star formation stops are the same as for our own Galaxy. What changes with
respect to the MW model is the 
value of the central mass surface density at the present time and the value 
of the disk scale-length 
in the disk exponential law ($\Sigma(R)$ = $\Sigma_0 \times \exp^{-R/R_D}$; 
here $R_D$ = 5.4 kpc rather than 3.5 kpc, which is the value
adopted for the MW). 

The value of $\Sigma_0$ is very uncertain. The disk surface
brightness at R$\simeq$2 kpc ($\simeq$ 1 arcmin) of M101 
in the B band is 21.45 mag/arcsec$^2$ according to Okamura et al. (1976).
For the solar neighborhood, we have a surface brightness of the order of 
23.8 mag/arcsec$^2$ (Van der Kruit 1986), 
which corresponds to a stellar density of $\sim$40 M$_{\odot}$/pc$^2$ (see CMR2001).
This means that the central surface brightness of M101 is about 2.5 magnitudes brighter
than the solar neighborhood, which is about a factor of 10. This would give 
a central surface mass density of around 400 M$_{\odot}$/pc$^2$ at around 2kpc, 
implying a $\Sigma_0$ $\simeq$ 600 M$_{\odot}$/pc$^2$. Given the uncertainties
in this value we run models for $\Sigma_0$ = 600, 800 and 1000 M$_{\odot}$/pc$^2.$
However, as we will see in the next section, this value does not affect the 
abundance ratios.

The observed H\,{\small I} and H$_2$ profiles are taken from Kenney et al. (1991), 
where a distance of 5.5 Mpc for M\,101 is assumed. 
The theoretical present-day gas density distribution critically depends on the 
assumptions on the gas density threshold ($\Sigma^{th}_{disk}$), 
time-scales of disk formation and 
central mass density. The gas density profile in the external parts of
M101 is practically determined solely by the choice of $\Sigma^{th}_{disk}$. 
On the other hand, the inner profile strongly depends on the assumed $\Sigma_0$ value and 
time-scale of thin-disk formation.

\section{Results}

\subsection{The Milky Way}

\subsubsection{The Solar and the Interstellar Medium Abundances }

The solar abundances should represent the chemical composition of the 
ISM in the solar neighbourhood at the time of Sun formation 
(4.5 Gyrs ago). The standard assumption is that since the time of the
Sun formation the ISM was further enriched
in several chemical elements. 
However, as often discussed in the literature,
the abundance of oxygen in the Orion nebula 
is smaller by a factor of 2 than the
solar value of Anders and Grevesse (1989) (see for example Cunha and Lambert 1992), 
at variance with the increase of the metal abundances in the 
Galaxy with time as predicted by many chemical evolution models. 
This fact, together with the abundances in other HII regions 
and B stars, led to the view that the Sun was enriched relative
to the ISM, a possibility being that the Sun was born in a region closer
to the Galactic center and then moved to its present position (Wielen et al. 1996).

Recently this situation has changed. The updated
solar O abundance by Holweger (2001) and
Allende Prieto et al. (2001) does not require to have a subsolar
standard abundance for the present time ISM. 
As can be seen in Tables 3 and 4, when considering the
revised photospheric oxygen abundance obtained by Holweger (2001) and 
comparing it with the Orion oxygen abundance (corrected for dust) 
given by Peimbert (1999), the ISM and 
Sun values are the same within the rms scatter. The same
is true for the young F- and G-type stars (Sofia \& Meyer 2001).
Moreover, the often reported discrepancy among the Sun and B-star abundances
could be due to the details of the evolution of the B stars themselves. In fact,
Sofia \& Meyer argue that B stars may not offer a good measure
of the ISM metallicity because processes of sedimentation and/or
ambipolar diffusion, occurring during their formation, could lower
their refractory abundances. This is confirmed by the recent
work by Daflon et al. (2001), where a sample of B stars with high rotational
velocities show a systematically lower abundance value when compared with
HII regions and with the predictions of chemical evolution models.
This does not seem to be the case for F- and G-stars which 
in principle are not greatly affected
by refractory depleting processes (Sofia \& Meyer 2001). 
 
The predicted abundances of C, N, O, Fe, Ne, S, Si and Mg 4.5 Gyrs ago
(to be compared with the solar values) and at the present time (to
be compared with recent data on the present abundances in the ISM)
are shown in Tables 3 and 4, respectively, for all the models of Tables 2a,b.

If the solar and the ISM values are similar, 
chemical evolution models should be able to 
explain this constraint.
The similarity between the solar abundances and the ISM ones could be 
indicating that the evolution of the solar vicinity in the last 4.5 Gyr was 
very slow. In our models (as well as in CMG97 and CMR2001), 
the last 4.5 Gyr of evolution of the solar vicinity are dominated by 
an oscillatory behavior of the star formation rate owing to the 
presence of the gas density threshold for the star formation. 
This is the case of all models of Tables 2a and 2b, except
for model 9. Model 9 was computed with the same parameters as 
model 6 but with the assumption of no threshold in the star formation
process in the disk. As can be seen in Table 3, models 6 and 9 give
the same results at the time of the Sun formation, whereas model 9
predicts larger abundances for the ISM at the present time (Table 4) compared
to the ones predicted by model 6. In particular, model 6 predicts only a small
increase of the elements produced by massive stars, from the time of Sun formation
up to now and is in agreement with the observed constancy of the oxygen 
abundance in the last 4.5 Gyr.

Figures 2 and 3 show the evolution in time of C, N, O, N/O and C/O for some
of the models of Table 2a. The solar values obtained by Holweger (2001),
Grevesse \& Sauval (1998), Grevesse
et al. (1996), Anders \& Grevesse (1989) and Allende Prieto et al. (2001, 2002) are also shown.
In Figure 2 (upper panel) we show those models in which the carbon yields in massive
stars were not modified (model 1: thin dotted line, model 2: thin long-dashed line,
model 3: thick dotted-line, model 4: thick solid line and model 5: thin solid line). 
As can be seen, all models predict
a solar value which is a little below the observed ones. Model 1 (with RV yields for the case $\alpha$=0) predicts a larger
carbon as in this case HBB is not operating.
However, this model predicts a too low nitrogen abundance (Figure 2 - middle panel), 
thus showing the 
necessity of considering some HBB in AGB stars. 
In Figure 2 (lower panel) we note that model 5 (thin solid
line) predicts a lower oxygen value, but still consistent with the recent
measurements of Holweger (2001) and Allende Prieto et al. (2001).

In Figures 3a and 3c we show again models 1 to 5 (curves are as in Figure 2). 
All these models, but model 1, reproduce
the solar N/O ratio, whereas almost all models predict a solar C/O ratio which is a little
lower than the observed one. In Figure 3b we show models 3a (dot-dashed line), 
6 (long-dashed thick line) and 7 (long-dashed thin line). 
Models 3a, 6 and 7 are analogous to models 3, 4 and 5, respectively, 
but with carbon yields in stars with masses in the range of 40-100
M$_{\odot}$ multiplied by a factor of 3 (notice that in the diagram N/O vs. O/H models
3, 4 and 5 are identical to models 3a, 6 and 7, respectively, as only the carbon
yields were modified). Model 7 is in good agreement
with the observations. This is due to the fact that
WW yields predict less oxygen than TNH ones  (see Figure 2 - lower panel). 
Model 6 is in reasonable agreement with the observed value if we consider the uncertainties
involved both in the data and in the stellar yields for C.
In fact, two large uncertainties play an important role in this plot: a) the solar value is very
uncertain for carbon (see the different solar measurements plotted in this figure - big
square, triangle, star and pentagon) and b) the yields of vdHG were computed with a large
$\eta_{AGB}$ parameter (they adopted, at solar metallicities, $\eta_{AGB}$=4 while in 
RV this value was $\eta_{AGB}$=0.33). 
If the mass loss in AGB is overestimated, the C
production is also underestimated. In Figure 3b this can be clearly seen as the middle curve
(model 6), computed with a variable $\eta_{AGB}$ parameter at low metallicities 
($\eta_{AGB}$=1 for Z=0.001 and $\eta_{AGB}$=2 for
Z=0.004), shows a larger carbon abundance in the first billion years of the evolution 
of the solar vicinity than model 3a, computed with $\eta_{AGB}=$4 at all metallicities.

In fact, as shown in Table 1, although the yields of vdHG for the case of variable $\eta_{AGB}$ 
predict a decrease in the integrated C yield as the metallicity increases, 
reaching a value lower than the one predicted by the models of RV with 
$\alpha=$1.5 at solar metallicity, the predicted
C yields at low metallicities are larger by a factor of $\simeq$10.
This suggests that $\alpha=$2, assumed in vdHG calculations, is too
large and that a lower value should be preferred (in agreement
with the result of Diaz \& Tosi 1986).

Finally, we notice that Mg seems to be underestimated by all
models (see Table 3). The fact that current stellar evolution models
for massive stars underestimate the Mg yields is a well known problem
(see Chiappini et al. 1999 and Thomas et al. 1998 for a discussion).

\begin{table}
\begin{center}
\small
\textsc{Table 3. Solar abundances by number in log(X/H)+12 ($^*$ at 4.5
Gyrs ago).}
\vspace{0.5cm}
\footnotesize
\label{}
\begin{tabular}{c c c c c c c c c}
\hline
Model & {C} & {N} & {O} & {Fe} & {Ne} & {S} & {Si} & {Mg} \\
\hline
1 & 8.41 & 7.40 & 8.80 & 7.46 & 7.74 & 6.99 & 7.46 & 7.27 \\
2 & 8.19 & 8.04 & 8.79 & 7.46 & 7.74 & 6.99 & 7.46 & 7.27 \\
3 & 8.18 & 7.96 & 8.77 & 7.46 & 7.74 & 7.00 & 7.47 & 7.28 \\
3a & 8.22 & 7.96 & 8.77 & 7.46 & 7.74 & 7.00 & 7.47 & 7.28 \\
4 & 8.22 & 7.97 & 8.77 & 7.46 & 7.74 & 7.00 & 7.47 & 7.28 \\
5 & 8.29 & 7.98 & 8.66 & 7.51 & 7.81 & 7.11 & 7.53 & 7.15 \\
6 & 8.26 & 7.98 & 8.77 & 7.46 & 7.74 & 7.00 & 7.47 & 7.28 \\
7 & 8.32 & 7.98 & 8.66 & 7.51 & 7.81 & 7.11 & 7.53 & 7.15 \\
8 & 8.26 & 8.06 & 8.77 & 7.46 & 7.74 & 7.00 & 7.47 & 7.28 \\
9& 8.26 & 7.97 & 8.77 & 7.46 & 7.74 & 7.00 & 7.47 & 7.28 \\
\hline
Data & & & & & & & & \\
\hline
H2001 & 8.59 $\pm$ 0.11 & 7.93 $\pm$ 0.11 & 8.74 $\pm$ 0.08 &  7.45
$\pm$ 0.08 & & & 7.54 $\pm$ 0.05 & 7.54 $\pm$ 0.06 \\
AP0102 & 8.39 $\pm$ 0.04 & & 8.69 $\pm$ 0.05 & & & & & \\
GS98 & 8.52 $\pm$ 0.06 & 7.92 $\pm$ 0.06 & 8.83 $\pm$ 0.06 & 7.50 $\pm$
0.05 & & 7.20 $\pm$ 0.06 & 7.56 $\pm$ 0.01 & 7.58 $\pm$ 0.06 \\
G96 & 8.55 $\pm$ 0.05 & 7.97 $\pm$ 0.07 & 8.87 $\pm$ 0.07 & 7.50 $\pm$
0.01 & & 7.20 $\pm$ 0.04 & 7.56 $\pm$ 0.01 & 7.58 $\pm$ 0.06 \\
AG89 & 8.56 $\pm$ 0.04 & 8.05 $\pm$ 0.04 & 8.93 $\pm$ 0.03 & 7.51 $\pm$
0.01 & [8.09] & 7.27 $\pm$ 0.05 & 7.55 $\pm$ 0.02 & 7.58 $\pm$ 0.02 \\
\hline
\end{tabular}
\end{center}
\footnotesize{H2001 - Holweger (2001), AP0102 - Allende Prieto et al. 2001, 2002, GS98 - Grevesse
\& Sauval 1998, G96 - Grevesse et al. 1996, AG89 - Anders \& Grevesse 1989}
\end{table}

\begin{table}
\begin{center}
\small
\textsc{Table 4. ISM abundances by number in log(X/H)+12 ($^*$ at t=t$_{now}$).}
\vspace{0.5cm}
\footnotesize
\label{}
\begin{tabular}{c c c c c c c c c}
\hline
Model & C & N & O & Fe & Ne & S & Si & Mg \\ 
\hline
1 & 8.48 & 7.58 & 8.87 & 7.61 & 7.76 & 7.09 & 7.53 & 7.30 \\
2 & 8.27 & 8.11 & 8.86 & 7.61 & 7.76 & 7.09 & 7.53 & 7.30 \\
3 & 8.27 & 8.05 & 8.81 & 7.62 & 7.78 & 7.10 & 7.55 & 7.31 \\
3a& 8.31 & 8.05 & 8.81 & 7.62 & 7.78 & 7.10 & 7.55 & 7.31 \\
4 & 8.29 & 8.06 & 8.81 & 7.62 & 7.78 & 7.10 & 7.55 & 7.31 \\
5 & 8.35 & 8.06 & 8.70 & 7.65 & 7.84 & 7.20 & 7.61 & 7.19 \\
6 & 8.33 & 8.06 & 8.81 & 7.62 & 7.78 & 7.10 & 7.55 & 7.31 \\
7 & 8.38 & 8.06 & 8.70 & 7.65 & 7.84 & 7.20 & 7.61 & 7.19 \\
8 & 8.32 & 8.14 & 8.81 & 7.62 & 7.78 & 7.10 & 7.55 & 7.31 \\
9& 8.34 & 8.09 & 8.85 & 7.63 & 7.81 & 7.12 & 7.57 & 7.35 \\
\hline
Data & & & &  &  & & & \\ 
\hline
FUSE-ISM (M2002) & & & 8.5-8.55 & & & & & \\ 
Young F \& G (SM01)  & 8.55 &  & 8.65 & 7.45 & & &  7.60  & 7.63 \cr
HST-ISM (M9798) &  & 7.88 & 8.50-8.70 & & & & & \\
Orion (P99) & 8.49 $\pm$ 0.12 & 7.78 $\pm$ 0.08 & 8.72 $\pm$ 0.07 &
7.48 $\pm$ 0.15 & & & 7.36 $\pm$ 0.20 & \cr 
\hline
\end{tabular}\\
\end{center}
\footnotesize{M2002 - Moos et al. 2002; SM01 - Sofia \& Meyer 2001; M9798 - Meyer et al. 1997, 1998; P99 - Peimbert 1999}
\end{table}

\subsubsection{The [O/Fe] versus [Fe/H] plot}

It is worth noting that the two-infall model provides a good fit of the 
[$\alpha$/Fe] versus [Fe/H] relation in the solar vicinity, as shown in 
Chiappini et al. (1999). Here we show the plot of [O/Fe] versus [Fe/H] 
which clearly indicates the existence of a gap in the SFR occurring at the 
end of the halo/thick-disk phase and before the thin-disk formation.
This gap is observed in the data as pointed out by Gratton et al. (2000). 
In fact, if there is a gap in the SFR we should 
expect both a steep increase of [Fe/O] at a fixed [O/H] and a lack of stars 
corresponding to the gap period (see discussion in CMR2001).
This gap, suggested also by the [Fe/Mg] vs. [Mg/H] diagram of Fuhrmann (1998), in 
our models is due to the adoption of the threshold in the star formation 
process coupled with the assumption of a slow infall for the formation of
the disk. 

Another interesting point concerning the abundance ratios that can be understood
using the particular case of [O/Fe] is discussed below. 
Israelian et al. (1998) performed a detailed
abundance analysis of 23 unevolved metal-poor 
stars and found that the [O/Fe]
ratio increases from 0.6 to 1.0 dex between [Fe/H] $-$1.5 and $-$3 dex. 
In that paper the abundances were determined using high-resolution OH bands in
the near UV (easier to observe
in low-metallicity stars). 
Their result led to a controversy, as
a linear run of [O/Fe] with metallicity 
is difficult to conciliate with the basic ideas on
the roles of type Ia and type II SNe 
in the ISM enrichment (Matteucci \& Greggio 1986;
Matteucci \& Chiappini 2001) as well as with measurements 
made by other groups (e.g, Fulbright \& Kraft 1999, Sneden \& Primas 2001, Nissen et al. 2001, 2002).

Mel\'endez et al. (2001) obtained high-resolution infrared spectra 
in H-band in order
to derive oxygen abundances from IR OH lines and found that for a sample
of stars in the $-$2.2 $<$ [Fe/H] $<$ $-$1.2 range, 
[O/Fe] $\simeq+$0.4 $\pm$ 0.2 dex
with no significant evidence for an increase of [O/Fe] with decreasing
metallicity. The same behavior
is seen in the compilation of Gratton et al. (2000), where O is measured 
from the [OI]$\lambda \lambda$ 6300 which is known to be the most reliable
way to estimate oxygen abundances (Lambert 2001).
Moreover, Asplund \& Garcia P\'erez (2001) 
show that traditional 1D LTE analyses of the UV OH lines 
can overestimate the [O/Fe] ratio and that when adopting
3D analyses the results from UV OH lines become consistent 
with the one obtained from oxygen forbidden lines. 
Note also that Nissen et al. (2002) predict
that derivations of oxygen abundance from [OI] 6300 line also should lower at low
metallicities in hot stars. The effects on cooler giants should be less important, but
no calculations are available at the moment.

In a more recent paper, Mel\'endez \& Barbuy (2002) publish new oxygen abundances
again obtained from infrared OH lines with 10m Keck Telescope. Moreover, Mel\'endez
\& Barbuy show a compilation of the best measurements (see details in section
4.2 of Mel\'endez \& Barbuy) so far published in the
literature. They
included only abundance data obtained from the oxygen [OI] forbidden line 
(Figure 4a - the stars show the data of Mel\'endez
\& Barbuy 2002 in the infrared, the dots show their 
compilation of oxygen data from the literature, where the size of the full dots represents
the number of stars in each bin of 0.2 dex in [Fe/H]). 

The data for oxygen, taken from Mel\'endez \& Barbuy (2002) (Figure 4a) 
show a slight increase of the [O/Fe] ratio with decreasing 
[Fe/H], at variance with what happens 
for other $\alpha$-elements which show a flatter behavior.
This slight slope is well reproduced by theoretical models 
(Chiappini et al. 1999, CMR2001 - models 4 and 5) owing to the fact that the O/Fe 
ejected by a massive star at the end of its evolution 
is an increasing function of the initial 
stellar mass (see figure 2b of CMR2001). 
As a consequence, even before the contribution
of SNe Ia to the iron enrichment, the O/Fe ratio already decreases
due to the different ejected quantities of O/Fe by massive stars of different
masses. The main change in slope occurs at [Fe/H]$\simeq-$0.6 dex and corresponds
to the time at which the type Ia SN rate in the disk reaches a maximum, roughly 1 Gyr after
the gap (see also Matteucci and Recchi 2001). 
 
On the other hand, our predictions for the [O/Fe] ratio, especially 
at very low metallicities, are not in agreement with the claims of 
a linear rising of this ratio with decreasing [Fe/H] obtained
from UV OH lines (Figure 4b). These measurements are systematically larger
than the samples shown in Figure 4a and are not reproduced by our model.

\subsubsection{The log(C/O) vs. log(O/H) plot}

By adopting the yields discussed before (with no mass loss
from massive stars), we can explain the increase of the C/O
abundance ratio as a function of metallicity as follows:
both $^{16}$O and $^{12}$C are primary elements, 
but $^{12}$C is mainly restored into the ISM with a time delay by 
intermediate-mass stars and hence on longer timescales compared
to the $^{16}$O enrichment, which comes mainly
from massive stars.

In Figure 5 we show our model predictions for log(C/O) vs.
log(O/H) and compare them with the available 
abundance data for the solar vicinity.
Carbon abundances can be obtained from the CH lines in the blue and 
ultraviolet (UV), infrared CO emission, red and near-infrared
CI lines and [CI] forbidden lines. The [CI] line is less dependent
on the effective temperature than the CI and CH lines. However,
the [CI] line can be very faint and difficult to measure at low 
metallicities.
In fact, both the [CI] and CI lines become extremely weak in metal-poor 
stars (already at [Fe/H] $\leq -$1) (see Carretta et al. 2000
for a detailed discussion on this element). Here we will consider
the more recent measurements of Mel\'endez et al. (2001) (small open squares),
Mel\'endez \& Barbuy (2002) (big open squares), the reanalysis of data by Edvardsson
et al. (1993) done by Carretta et al (2000) (open circles) and the data of 
Carretta et al. (2000) (oxygen values obtained 
as the mean of the abundances obtained with [OI] and OI oxygen lines - open triangles - from only 
[OI] measurements - filled triangles). 
The lack of data at low metallicities is striking. 
Moreover, we can see a large spread at low metallicities.
Some of the stars at low metallicities show enhanced C abundances due to 
internal mixing processes. To avoid these objects
we plot in Figures 5c and 5d only dwarf stars
(i.e., objects with $\log g >$ 3.5).

Panels a) and c) show our model predictions 
for models 1 to 5, while in panels b) and d) we show the predictions of 
models 3a, 6 and 7 (see Table 2a).
The first thing to
notice is that the effect of the gap in the star formation
is also seen in this plot. 
The confirmation
of a gap in the star formation in such a plot 
would be another evidence in favor
of the interpretation that most of the C we see today
in the ISM comes from low- and intermediate-mass stars. In fact, if C were
produced mainly in massive stars, as suggested by Prantzos et al. (1994), 
we would not expect to see any gap in such a plot, 
as both O and C would depend ``linearly''
on the SFR.
This figure also shows that 
models adopting WW yields predict a larger
C/O ratio (as WW models have more C and less O than
the TNH ones - see Table 1).

We expect the effect of considering stellar yields
including mass loss in massive stars in chemical evolution
models to be mild. According to stellar models with mass loss, 
the C enrichment of the ISM from massive stars starts to be important
only at large metallicity regimes. However, at this point, low-
and intermediate-mass stars also contribute to the ISM enrichment
in C and, as they are more numerous, their contribution
will certainly be dominant.
Moreover, as discussed before,
the new calculations of Meynet \& Maeder (2002b) point to a less important
effect due to mass loss. Note, however, that rotation could increase
the C yields which would shift our curves upwards in Figure 5. At this point
we call attention to the fact that it is not correct to plot 
the integrated stellar yields
on diagrams of this kind (as done in Meynet \& Maeder 2002b - their Figures 20 and 21)
as this could lead to wrong interpretations. Chemical
evolution models which take into account properly the stellar lifetimes have
to be used. This is of fundamental importance in order to understand which are 
the stars playing an important role in the ISM enrichment at different metallicities.

Finally, in Table 3 it can be
seen that models for which the C yields are artificially increased (e.g. model 7)
are in better agreement with the solar abundance.
The need for increasing the C yields in massive stars
was already noticed by Henry et al. (2000), who
increased the yields by the same factor.

\subsubsection{The [C/O] vs. [Fe/H] and [C/Fe] vs. [Fe/H] plots}

The study of the evolution of C/O cannot be disentangled from the analysis of 
the [C/Fe] vs. [Fe/H] diagram since this latter represents an equally important
and independent constraint on the evolution of the C abundance.

Figure 6a shows the plot of [C/Fe] vs. [Fe/H] for 
models 1 to 5 (Table 2a) compared with the data from 
the compilation of Chiappini et al. (1999) (open squares).
The observations imply an essentially solar [C/Fe] ratio along the whole
metallicity range in agreement with our model predictions.
Figure 6b includes only the most
recent abundance data by Carretta et al. (2000), Mel\'endez et al. (2001),
and Mel\'endez \& Barbuy (2002). In this
case the data seem to show [C/Fe] $<$ 0 at low metallicities, exactly
the opposite of what is suggested in Figure 6a. However, when 
only dwarf stars are plotted (Figure 6c), very few data points are left at
low metallicities and most of the stars with low [C/Fe] ratios
disappear (in this figure models 3a, 6 and 7 are shown). These stars probably suffered  
HBB and part of their carbon was burned into nitrogen, and hence
their chemical composition cannot be used as a tracer of the ISM composition
at the time the star was formed. 
If the trend shown in Figure 6 is confirmed and [C/Fe]$\simeq$0
for all metallicities, this indicates that C, as Fe, is
ejected into the ISM on long timescales (at variance with the results of, for 
example, Carigi 2000, Henry et al. 2000, Liang et al. 2001).
To better constrain the C production in massive
stars, more data on C in dwarf halo stars are of fundamental importance.

Figures 7a,b show the plot of [C/O] vs. [Fe/H]. In figure 7a we 
show the model predictions for models 1 to 5 listed in Table 2a.
In this plot, as we did in Figure 6b, we also show only the best 
measurements for carbon (i.e., Mel\'endez et al. 2001; Carretta et al. 2000; 
and the reanalysis by Carretta et al. 2000 of the data of 
Edvardsson et al. 1993).
Again, a large scatter is seen at low metallicities, which makes
it difficult to constrain our models. 
In figure 7b only models 3a, 6 and 7 are shown, and only the dwarf stars are plotted.
As can be seen in this figure, both models are in good 
agreement with the available data.

\subsubsection{The log(N/O) vs. log(O/H) plot}

In Figures 8a,b we plot log(N/O) as a function of log(O/H).
The predictions of models 1 to 5 are shown. 
The lack of data in these plots is due to the fact that
the abundances of N and O are measured simultaneously only
for a few stars.
We also included the 
data by Daflon et al. (2001) for B stars (stars).
As discussed
before, the B stars lie systematically below the predicted curves
and the solar values, suggesting that processes like sedimentation
and/or ambipolar diffusion took place in such stars (see 
Sofia \& Meyer 2001 and discussion in Section 4.1.1). 
The solar abundances obtained from different
authors are also shown (large square, triangle, star and hexagon).
In Figure 8b, where we plot only dwarf stars, the data points at low 
metallicity disappear. In fact, those stars show both high N/O and 
low C/O ratios (see discussion of Figure 6 in the previous section), 
which suggest that they have suffered mixing 
processes during their evolution and are not suitable for this 
kind of analysis. In other words, the C and N values
in giants are not representative of the halo composition at the time
of their formation.
 
Figure 8 shows clearly the secondary behavior of N at low metallicities: the N/O ratio
increases with oxygen until it reaches a plateau due to the primary N restored
by low- and intermediate-mass stars. 
Finally, an interesting aspect of this plot is the existence
of some stars with low N/O which lie
near the prediction of model 1, computed with the 
stellar yields of RV for the case of no HBB ($\alpha =$0). 
One possibility is that these stars did not suffer HBB, suggesting
that HBB is not efficient in all intermediate-mass stars.
It would be interesting to study the properties of these particular
data points in order to better constrain both the lower-limit mass
for HBB to operate and the HBB efficiency.

\subsubsection{The [N/Fe] vs. [Fe/H] plot}

Figure 9a shows the [N/Fe] vs. [Fe/H] plot. The open squares
are from the sample studied in Chiappini et al. (1999).
The data of Carretta et al. (2000) are also plotted (here we
show only their measurements for dwarf stars - filled
squares).
Again a large scatter is seen in the data. Most of the points
at low metallicities are from Laird (1985) and Carbon et al.
(1987). These data sets were further corrected
and offset to take into account errors in the temperature. Here we 
do not apply such a correction which was very uncertain (see Wheeler
et al. 1989 for a discussion). Note that the effects of 3D hydrodynamical
models would also lower the N derived from NH bands and, in principle, most 
of the data available in the literature should be considered as upper limits
to the real N abundance value (Nissen, 2002).

From this figure it can be seen that N shows again the behaviour of a secondary element, 
especially at low metallicities (models 1 to 5). 
This fact should argue in favor of a secondary N
production in massive stars.
The [N/Fe] ratio increases with metallicity up to [Fe/H]$\sim-$1
and then flattens for larger metallicities in analogy with what happens
from N/O. Therefore, by analyzing only the 
data available for the MW there is no need for invoking
a primary N production in massive stars. 
In Figure 9b we show the prediction of model 8
where we adopt a large production of primary N in 
massive stars (see Table 2b). 
We can see that in order to be
able to have a plateau in [N/Fe], as often suggested in the literature,
a large amount of primary N from massive stars is needed (in model 8 -- dash-dotted line -- we assume a primary N yield from massive stars of 
0.065 M$_{\odot}$, constant with mass as suggested by Matteucci 1986).
This amount is larger by roughly two orders of magnitude than the one predicted by recent
theoretical models suggesting the existence of some primary
N production in massive stars at low metallicities (Meynet \& Maeder 2002a).

In summary: {\it i)}
our results indicate that the N yields computed by vdHG predict too 
much primary N from intermediate mass stars at low
metallicities. In fact, these yields were
computed for a case of maximum HBB efficiency; 
{\it ii)} the MW data do not suggest the need of primary N production 
in massive stars. We also show that 
to obtain a flat behaviour of [N/Fe] vs. [Fe/H] the requested quantity
of primary N from massive stars is much larger than the
one computed by recent models. Our conclusions should hold also 
for all other kind of galaxies, as there
is no reason for the nucleosynthesis to change.
In Section 4.2 we will check the validity of this statement
by analyzing other galaxies; {\it iii)} the flat behavior of [C/Fe] vs
[Fe/H] clearly indicates that both C and Fe should come mainly from
low- and intermediate-mass stars; {\it iv)} C yields in low- and
intermediate-mass stars are underestimated at solar metallicities probably
due to the large value of $\eta_{AGB}$ adopted by vdHG. Moreover, our models suggest
that the C yields in massive stars
(TNH and WW) should be increased by at least a factor of 3.
This is in agreement with recent stellar evolution models taking into account rotation (Meynet \&
Maeder 2002b).

\subsubsection{Abundance gradients of C, N and O in the Milky Way disk}

Our prescriptions for the MW model outside the solar vicinity are the same
as model A of CMR2001 (see also Section 3.1). 
Figure 10a shows the gradients of C, N and
O predicted at the present time and compared with HII regions and B star
data (see CMR2001 for details). Figure 10b shows the gradients of C/O,
N/O and C/N, both predicted and observed. 
In those figures we plot models
3a (dash-dotted line), 6 (long-dashed thick line) and 7 (long-dashed thin line).
The difference among models 6 and 7 is only on the stellar yields adopted
for massive stars (TNH and WW, respectively), whereas models 6 and 3a differ
on the adopted yields for low- and intermediate-mass stars (model 6 has
$\eta_{AGB}$ variable with metallicity, while for model 3a $\eta_{AGB}$ is
constant).

As can be seen in Figure 10a and 10b (upper panels), for C very few data are available 
(in Figure 10b the open triangles are the HII regions measured by Esteban et al.
1999, the open circles are the HII regions of Tsamis et al. 2002 and 
the black triangles are the data from Gummersbach
et al. 1998 for B stars). B stars lie systematically
below the HII regions. 

As shown by Figure 10b (upper panel), models 6 and 7 predict the same behavior 
for the abundance gradient of C/O. 
The absolute value of the C/O abundance ratio is larger for model
7 (WW yields) than for model 6 (TNH yields). This is expected as WW yields
produce more carbon and less oxygen than TNH ones (see Table 1). 
A more interesting effect can be seen when comparing the predictions of models
6 and 3a. These models differ only at low metallicities (where $\eta_{AGB}$=1
for Z=0.001 and $\eta_{AGB}$=2 for Z=0.004 for model 6, whereas for model 3a
at these same metallicities $\eta_{AGB}$=4) and hence differences should
be seen only in the outer parts of the Galactic disk. 
As discussed before, lower $\eta_{AGB}$ 
values imply a larger production of carbon and this explains the larger 
C/O ratios obtained with model 6. Model 3a, instead, predicts a 
flatter behavior of the C/O ratio along the disk. Unfortunately abundance data 
are not available in the galactocentric distance range where the models
differ. 

The predicted N/O gradient
for the MW is almost flat (Figure 10b - middle panel) 
and in good agreement with the observations.
In the case of N, the stars which contribute most to the ISM 
enrichment at low metallicity are those suffering HBB (i.e., stars with masses larger 
than 4-5 $M_{\odot}$). These intermediate-mass stars start
to contribute with a large amount of primary N already after 30 Myrs
(the lifetime of an 8$M_{\odot}$ star). 
In other words, the expected difference on the timescales of the enrichment
of the ISM in N and O is small.
This explains why the N/O ratio shows a flatter gradient than the C/O ratio.
Finally, Figure 10b (lower panel) shows the predicted gradients of C/N.
The flat C/N gradient predicted by model 3a is due to the fact that this model
predicts less C in the outer parts of the Galactic disk (lower metallicities)
than models 6 and 7.

\subsection{External Galaxies}

In this section we analyze the CNO data in other
galaxies by using the chemical evolution models described in Section 3.
For all the models discussed here the nucleosynthesis adopted is
the same as model 6 (TNH yields for massive stars and vdHG for the 
case of $\eta_{AGB}$ variable). 

\subsubsection{Results for M101}

For the Milky Way there is a clear lack of abundance data beyond $\simeq$
10-12 kpc from the Galactic center. 
In the case of external spiral galaxies it is possible to overcome this 
problem and measure the abundances of the far outermost HII regions, but
there are still uncertainties. One of the main difficulties in measuring 
oxygen abundances in the outermost HII regions of spiral galaxies
is to observe the [OIII] 4363 line, which is often not detected and which
is very important to estimate the oxygen abundance. In most of
the cases the oxygen abundance is rather estimated from the R$_{23}$ calibration
(Pagel et al. 1978; Edmunds \& Pagel 1984). In this latter case, the degeneracy 
between the upper and lower branches of the log(O/H) vs. R$_{23}$ relation
has to be resolved by using other
line ratios (see van Zee et al. 1998a; Ferguson et al. 1998). It is 
important then to keep in mind that for the outermost HII regions
the oxygen abundance could be very uncertain (see discussion in van Zee
et al. 1997).

To model the M101 galaxy, we adopt different values for the central total surface
mass density, namely $\Sigma_0^{M101}$ = 
1000, 800 and 600 $M_\odot$ pc$^{-2}$ together with a value 
of $R_D$ = 5.4 kpc for the disk scale-length (van Zee et al. 1998a). 
These are the fundamental
parameters for this model and can in principle be inferred from
observations, although the value of $\Sigma_0^{M101}$ is very uncertain
(Section 3.3). We notice that when adopting the scaling law suggested
by Boissier and Prantzos (2000), we get a value for the central
density of the M101 disk which is too low (in fact, this scaling law
suggests that a more massive galaxy as M101 would be less
concentrated and hence will have a lower central surface
mass density value). Models with $\Sigma_0^{M101} < \Sigma_0^{MW}$
could not reproduce the observed total gas density profile even
when allowing for different prescriptions for the SFR, IMF, etc.
In summary, our first result with respect to M101 is that its central
surface mass density should be larger than the one for the MW.

Once these parameters are fixed, the main
free parameters still left are the infall timescales for the disk formation and
their variation with galactocentric distance. These can be constrained
by both the gas profile and the abundance gradients.
From the study of the MW we know 
that the disk was formed inside-out and that the timescale
for the disk formation is an increasing function of Galactocentric
radius (CMR2001). 
Moreover, as shown by abundance data in other spiral galaxies,
more massive galaxies show flatter gradients (Zaritsky et al. 1994) 
and have higher average interstellar abundances (Garnett et al. 1997a),
thus suggesting that the more massive is the galaxy
the faster is the collapse of the gas forming the disk. 
This means that the variation of the accretion timescale of gas onto the disk
should be also a function of the galaxy mass (see Boissier \& Prantzos 2000),
in the sense that the most massive galaxies formed in the shortest timescale.

Figure 11 shows the observed gas profile for M101 (the total 
gas profile is shown by the short-dashed line - Kenney et al. 1991). Our predictions
are shown for the models adopting different values of $\Sigma_0^{M101}$
(solid line for $\Sigma_0^{M101}$=1000; dotted line for $\Sigma_0^{M101}$=800
and long-dashed line for $\Sigma_0^{M101}$=600).
Although the gas profile depends upon several model parameters, it is more
sensitive to the 
adopted total mass density profile and to the infall timescale
of gas accretion onto the disk. 
The adopted $\tau_D(R)$ values in this case 
are systematically lower than the ones of the MW, and the difference
between the internal and external $\tau_D(R)$ is smaller. 
In particular, we adopted $\tau_D(R) \sim 0.75 R - 0.5$, where
R is given in kpc and $\tau_D(R)$ in Gyr.
Since we assumed the same $\tau_D(R)$ for models with different values
of $\Sigma_0^{M101}$, all of them give roughly the same values for the abundance
ratios. 
The $\tau_D(R)$ law for M101 was chosen
to reproduce the abundance gradient of oxygen and to
ensure a good fit to the total gas mass profile.

Figure 12 (upper panel) shows the gradient of oxygen
measured by van Zee et al. (1998a), Kennicutt and Garnett (1996) and Garnett et al. (1999) 
together with our model predictions (here
we show only the model with $\Sigma_0^{M101}$=1000).
The model leads to a predicted oxygen gradient which
is flatter than the observed one in the outer parts. The short-dashed line refers to essentially
the same model, but here we adopted a larger threshold value (of 15 $M_{\odot}/pc^2$)
in the outer region (R$\geq$ 14 kpc). This last model provides
a better fit to the observed oxygen abundance gradient being still
in agreement with the observed gas profile.
Figure 12 also shows the C/O and N/O gradients for M101 (middle and lower panel, respectively). 
As can be seen in this figure, our model is in good agreement with the 
C/O abundance ratio measured in the outer parts of M101 (at R$\sim$15 kpc - filled square), 
while it probably predicts lower values than observed in the inner parts (there is only
one measurement at R$\sim$10 kpc for which the two values given by Garnett et al. 1999,
depending on the adopted reddening correction, differ by more than .2 dex). 
This is again showing what 
we already discussed when analyzing the abundance data in the MW namely, that
the C yields at solar metallicities are underestimated probably owing to an overestimated
value of $\eta_{AGB}$. 
In the lower panel of Figure 12 the opposite happens: 
the model predicts too large values of N/O
in the outer parts and ensures a good fit of the abundance data in the inner parts.
This last result suggests again that the primary N 
in intermediate-mass stars computed by vdHG is overestimated (see also Diaz \& Tosi 1986).

Moreover, we notice that the outermost point 
of Garnett et al. (1999) is around 4.5 scale lengths
from the center (R$\sim$25 kpc). At such large distances the threshold in the star
formation should be a dominant process, and our models predict that almost no
star formation took place,
and that low values for C, N and O abundances were produced.
This would explain the low N/O, C/O and O/H values measured in the outer
parts of M101. 

In summary, the results for M101 suggest that:
{\it i)} the nucleosynthesis adopted for the MW is suitable also for
other spiral galaxies
(we show here the particular case of M101 for which the gradients
of both N/O and C/O are available); {\it ii)} also in this
case there is an indication that the primary N coming from intermediate-mass
stars could have been overestimated
in the stellar models of vHG; {\it iii)} primary N production in massive stars 
would worsen our predicted N/O gradient which is already flatter than the observed 
one;
{\it iv)} the threshold gas density in the star formation can
explain the low abundance ratios measured in the outer parts of spiral galaxies.
A threshold gas density value which increases towards the outer parts of spiral disks
is suggested by our results.

\begin{deluxetable}{ccccc}
\small
\tablecolumns{5}
\renewcommand{\tabcolsep}{4pt}
\renewcommand{\arraystretch}{.6}
\tablewidth{0pt}
\tablenum{5}
\tablecaption{\small{Dwarf Galaxies Models}}
\tablehead{
\colhead{Model} & \colhead{number of bursts} & 
\colhead{time of burst occurrence} & \colhead{burst duration} & \colhead{$\nu$} }
\startdata
A$^*$ & 1 & 13.98 & 0.01 & 1.0\\
B$^*$ & 1 & 13.98 & 0.01 & 2.5\\
C & 1 & 6.0 & 0.01 & 2.5\\
D$^*$ & 1 & 6.0 & 0.01 & 1.0\\
E & 1 & 6.0 & 0.5 & 2.5\\
F & 1 & 6.0 & 0.5 & 1.0\\
G & 1 & 10.0 & 0.01 & 2.5\\
H & 2 & 6.0/13.98 & 0.01 & 2.5\\
I & 2 & 6.0/13.5 & 0.05/0.01 & 2.5\\
J & 2 & 6.0/13.98 & 0.1/0.01 & 2.5\\
K & 2 & 6.0/13.98 & 0.5/0.01 & 2.5\\
L & 2 & 13.85/13.95 & 0.05 & 2.5\\
\enddata
\footnotesize{$^*$ Models that do not develop galactic winds}
\end{deluxetable}

\subsubsection{Results for dwarf galaxies}

Carbon, nitrogen and oxygen data are available for dwarf irregular galaxies, 
also called extragalactic HII regions. These objects are interesting to study
since they are relatively simple objects with low metallicities and high gas
content. In particular, they are useful to study the abundances and abundance
ratios at low metallicities. In the present work we consider the
CNO abundance data from several sources (Garnett et al. 1995a,b; Garnett et al. 1997b;
Izotov \& Thuan 1999; Kobulnicky \& Skillman 1996; van Zee et al. 1997; Kennicutt \& Skillman 2001;
Thuan et al. 2002). In Figures 13a and 13b we show the available data
for N/O and C/O as a function of log(O/H), respectively.

The models for spiral galaxies do not apply in this case. 
As already extensively discussed in the literature, dwarf galaxies 
should have a bursting mode of star formation and they are likely to develop galactic
winds. Moreover, infall of primordial matter is likely to happen
in most of the cases as suggested by the large HI halos surrounding many of 
these objects. The main characteristics of these objects are 
the spread observed in their physical
properties (Matteucci \& Chiosi 1983).

For the dwarf irregular galaxies we adopted the chemical 
evolution model described in Section 3.2.
A more detailed discussion about this kind of galaxies can be
found in Recchi et al. (2001, 2002).
The main free parameters in our models are shown in Table 5:
the number of bursts, the time of occurrence
of the bursts (in Gyr), the duration of the bursts (in Gyr) and the efficiency
of the star formation (which also regulates the efficiency of
galactic winds - in Gyr$^{-1}$).
We run several models which are represented in Figures 13a,b by letters.
Some of the models in Table 5 do not develop galactic winds (namely, models A, B 
and D). 

In all those models, the nucleosynthesis adopted is 
the same as in model 6 for the MW.
In Figures 13a,b we show only the ``end point'' of each track and not
the time evolution of the abundance ratios. The reason for this is that in these figures
each observational point represents a different galaxy 
as seen at the present time and there is no information about
time evolution. In other words, this diagram is not the equivalent of the 
N/O vs. O/H for the solar neighborhood stars where O/H can be interpreted
as a time axis.

From Figure 13a it can be seen that the data for N/O show a flat trend 
with metallicity for a limited subsample of
blue compact dwarf galaxies, whereas in galaxies with log(O/H)+12 $>$ 7.6 there is a large
scatter. This has been often quoted in the literature as a reason to believe
that N could have a primary origin in massive stars (e.g., Izotov \& Thuan 1999, 
Meynet \& Maeder 2002b, and references therein). However, as it will be shown below,
chemical evolution models of dwarf galaxies can explain 
the ``high'' N/O ratios measured in low-metallicity objects without
the necessity to assume primary N contribution from massive stars (see also Henry et al. 2000).
Moreover, the low number of objects with log(O/H)+12 $<$ 7.6 so far observed challenges
the idea of a ``constant plateau'' at low metallicities. 

Our models indicate that objects suffering an initial
burst in the past plus a more recent one, with a low star 
formation efficiency or, alternatively, just one burst whose age is 
long enough that intermediate mass stars have had time to restore their N into the ISM, tend
to cluster around the N/O plateau (e.g. model C). 
On the contrary, the high metallicity
and high N/O part of the plot tends to be populated by models in which more than one
burst has occurred (e.g., model K). Moreover, the longer is the burst
duration the larger is the final N/O abundance obtained (Models E, F and K). Those models are also
in good agreement with the data in the log(C/O) vs. log(O/H) plot. We notice
that one of the main requirements to fit both the C/O and N/O abundance 
ratios is that those galaxies are characterized by an IMF flatter than
the Salpeter one. In fact, if a normal IMF is adopted, most of the
models tend to predict too high N/O ratios. 

Our main conclusions with respect to dwarf galaxies are summarized below. 
The large spread observed both in the N/O vs. O/H and
C/O vs. O/H diagrams for oxygen abundances larger than 7.6 can be explained
as due to different chemical evolution histories of different galaxies (i.e.,
different star formation efficiencies, different burst ages and 
different burst durations). Moreover, the time-delay in restoring N relative to O
into the ISM is also a cause for the observed spread (Larsen et al. 2001).
Once again we do not see the need
for invoking primary N production in massive stars to explain the above data. 
The adopted yields from vdHG already predict a large primary N contribution
from AGB stars. 
Objects with low oxygen
abundances and high N/O ratios could have simply suffered a burst at least
30 Myrs ago (the lifetime of an 8M$_{\odot}$ star), with a low star formation 
efficiency. Otherwise, they could be objects which had a previous burst of
star formation, a long quiescent period and start now to have a second burst.

\subsubsection{Comparison with observations in DLAs}

In this section we compare our model predictions for two
different regions of the MW and for dwarf irregulars with the abundance data
of DLAs. In this paper we discuss only the N/O and N/S
abundance ratios in DLAs. 
For a detailed discussion on all the other elements
usually measured see Calura et al. (2002). 

In Figure 14 we plot the recent data compilation by Pettini et al. (2002, 1995).
Pettini et al. (2002) include DLAs which have relatively low values of 
hydrogen column density in order to increase the probability that the 
O I $\lambda$1302.2 line is not strongly saturated (a known difficulty in this
kind of analysis) and which are at relatively low redshifts ($z_{abs} < 2.7$), where
the Lyman $\alpha$ line can still be clearly observed. 
The full sample described in Pettini et al. (2002) consists of 15 absorption systems in which
the abundance of N has been measured (11 detections and 4 upper limits).
In 8 of the 15 cases the abundance of oxygen is available directly, while in the 
other 7 cases S was measured and used as a proxy for oxygen (see also Molaro 2002).

In this figure we plot both N/O and N/S as a function of log(O/H)+12 
and log(S/H)+12, respectively. The open symbols refer
to oxygen while the filled symbols to sulphur data.
The Sun location in terms of S and
O is also shown. We also plot the points given by Lu et al. (1998).
The solid curve is the model 6 prediction for the MW in
the solar vicinity, for N/S (Figure 14, upper left) and N/O 
(Figure 14, lower right) as a function of S and
O, respectively. The dotted curve is the model 6 prediction
for an outer region (here we take the Galactocentric 
distance R $=$18 kpc). The letters A, C, H and J correspond to 
different models for dwarf
galaxies (see Table 5).

Two important things can be noticed in this figure.
The first one is that most of the N/O data fall below the 
so called ``N plateau'' defined by 
dwarf galaxies (log(N/O)$\simeq-$1.5). 
These data show lower N/O ratios, favoring the
time-delay model of N production from low- and intermediate-mass 
stars. This is in agreement with our previous
conclusion that primary N production in massive stars should 
not be important. 

The second thing is that our model predictions fit both the
N/S and N/O ratios in the sense that our curves for the MW 
reach log(N/O)$\simeq -$2 and log(N/S)$\simeq 0$ after roughly 100 Myr 
after the SFR started.
In particular, our model prediction for log(N/O) vs. log(O/H)
for a galactocentric distance of 18 kpc is in agreement with the
observations of Pettini et al. (2002) for oxygen (open squares in Figure 14).
This suggests that those objects could be similar to the outer parts of the MW
at early times (but see below). Notice that the often 
adopted procedure in the literature of converting the
N/S observations into N/O values by adopting a solar log(S/O)
would lead to a large scatter in the log(N/O) vs. log(O/H) plot.
In fact, when computing a detailed chemical evolution model
we see that S/O is not constant, especially at
low metallicities. As in the case of O/Fe previously 
discussed (see Section 4.1.2), the S/O ratio varies with metallicity
because {\it i)} the S/O ejected by 
massive stars varies with the initial mass of the
progenitor star and {\it ii)}
type Ia SNe contribute a non-negligible amount of S. 
Moreover, although the bulk of primary N contribution
by low- and intermediate-mass stars takes 300 Myrs, an important
amount of N is released into the ISM already after 30 Myrs (by
the 8 $M_{\odot}$ stars). This could explain part of the scatter
seen in the N/O ratios in DLAs. 

These results suggest that DLAs have experienced little
star formation up to the time they are observed. This is 
in agreement with the idea that a large fraction of 
such systems are either dwarf irregular galaxies or 
external parts of disk galaxies (see also Calura et al. 2002). 
In fact, there are essentially 4 ways of obtaining a low N/O at low metallicities:
a) with an IMF which is flatter than a Salpeter one plus an
inefficient star formation rate; b) that these systems are similar to outer
disks where the star formation was essentially zero for most of their evolution
due to the threshold and then started forming stars not before $\sim$100 Myrs
(depending on the assumed IMF and stellar yields) since the time at which they are observed.
In fact, in this case AGB stars have not had yet the time to
restore their primary N into the ISM; c) that these objects are precursors
of dwarf galaxies in which selective outflows
took place, ejecting more N than O and thus lowering their N/O ratios or 
d) that DLAs are precursors of dwarf galaxies observed either in the
burst or inter-burst phases. In this case the N/O ratio during the burst phase
is lower than in the inter-burst phase, where only N is produced.
However, the case of a very flat IMF can be excluded by the observed low $\alpha$/Fe.

\section{CONCLUSIONS}

We studied the evolution of the C, N, O and Fe abundances 
as a function of time and galactocentric
distance $R$ in the Milky Way.
In particular we computed the evolution of the C/O and N/O ratios,
and obtained the abundance gradients of C/O, N/O and C/N 
in the disk by means of a chemical 
evolution model which reproduces 
the majority of the observational constraints in the Galaxy (CMG97, CMR2001).

From the comparison between observations and model results for 
the Milky Way we can draw the following conclusions:

\begin{itemize}

\item
Models for the MW adopting 
van den Hoek and Groenewegen (1997) yields for low- and 
intermediate-mass stars and Thielemann et al. (1996) for massive stars are
well in agreement with the abundance data on metal-poor
stars, in particular the data for [O/Fe] obtained from
[OI] and IR OH lines;

\item
We show that [C/Fe]$\sim$0 over the whole [Fe/H] range clearly
indicates that both C and Fe should come mainly from low- and intermediate-mass stars. 
This is at variance with the interpretation by several authors
(e.g., Carigi 2000, Henry et al. 2000) that C should originate mainly in massive stars.
This conclusion was based on the yields of Maeder (1992) which overestimate the 
effects of mass loss in massive stars. As shown by our good fit of [O/Fe] vs. [Fe/H],
the yields of TNH are in good agreement with the observations, except for the fact
that their C values should be increased by a factor of 3 for stars with M$>$40M$_{\odot}$.
This is in agreement with the new calculations of Meynet \& Maeder (2002) taking into 
account rotation.

\item
A gap in the star formation rate between the thick and thin disk formation 
affects our model predictions for the C/O and/or N/O versus O/H plot. The
existence of such a gap is already confirmed in [Fe/O] vs. [O/H] and [Fe/Mg] vs. [Mg/H]
by observations (Gratton et al. 2000, Fuhrmann 1998). For the C/O and N/O versus
O/H plot it is not possible to conclude the same from the few/uncertain available data.

\item The threshold in the star formation 
rate is responsible for the slow chemical enrichment of the solar 
neighborhood in the last 4.5 Gyrs which explains the similarity between 
the observed abundances in Orion and the Sun.

\item The [N/Fe] vs. [Fe/H] plot for halo stars show that N has a clearly secondary
behavior at low metallicities thus implying that primary N production in massive stars
is not important. Moreover, to obtain a solar [N/Fe] along the whole
metallicity range, the required primary
N is at least two orders of magnitudes higher than that predicted by the most detailed
and up to date stellar evolution models (Meynet \& Maeder 2002a). 

\item Our model predictions for the abundance gradients of C, N and O are in good agreement
with the observations. In particular, we obtain a gradient for N/O which is 
flatter than the one of C/O. This is mainly due to the important contribution
of primary nitrogen from intermediate-mass stars suffering HBB. More data are necessary
to better constrain the C/O gradient in the Galaxy.

\end{itemize}

When applying the ``best'' nucleosynthesis prescriptions to other galaxies
we find that: 

\begin{itemize}

\item
Again, there is no need for claiming the existence of an important
primary N contribution from massive stars to explain the 
abundance data of dwarf galaxies, outer spiral disks or DLAs;

\item
Models computed with TNH and vdHG yields can well reproduce the distribution
of dwarf galaxies in plots such as log(N/O) and log(C/O) versus log(O/H)+12;

\item
The nucleosynthesis adopted for the MW is also
in agreement with what is observed in other spiral galaxies
(we show here the particular case of M101 for which the gradients
of both N/O and C/O are available). As for the MW,
there is an indication that the primary N coming from intermediate-mass
stars is overestimated and that C is
underestimated at solar metallicities; 

\item
Invoking primary N in massive stars does not solve the problem of explaining
the observed N/O gradient in M101. Instead, we suggest that the 
threshold in the star formation rate could 
explain the low abundance ratios at large galactocentric distances.

\item Our predictions for the abundance gradients
of N/O, C/O and C/N are also compatible 
with the idea that most of the C and N we see today in the ISM
comes from low- and intermediate-mass stars. 

\item Our model for M101 suggests that the threshold in the gas density
should increase with galactocentric distance.

\end{itemize}

Finally, we stress that the N/O vs. O/H diagram for dwarf galaxies
is often interpreted in the literature (the same is true
for the C/O vs. O/H diagram) as an evolutionary diagram, but instead
it represents the final abundance values achieved by
objects which evolved in a completely different way from each other.
In other words, this diagram is not the equivalent of the 
N/O vs. O/H for the solar neighborhood stars which is a 
real evolutionary plot, where for each O/H value corresponds a different
galactic age.
Therefore, any conclusion on the secondary/primary value of N derived
from such a diagram should be taken with care.

\begin{acknowledgements}
We acknowledge S. Recchi and M. Tosi for interesting discussions. 

\end{acknowledgements}

{}

\newpage

\begin{figure}
\figurenum{1}
\centerline{\psfig{figure=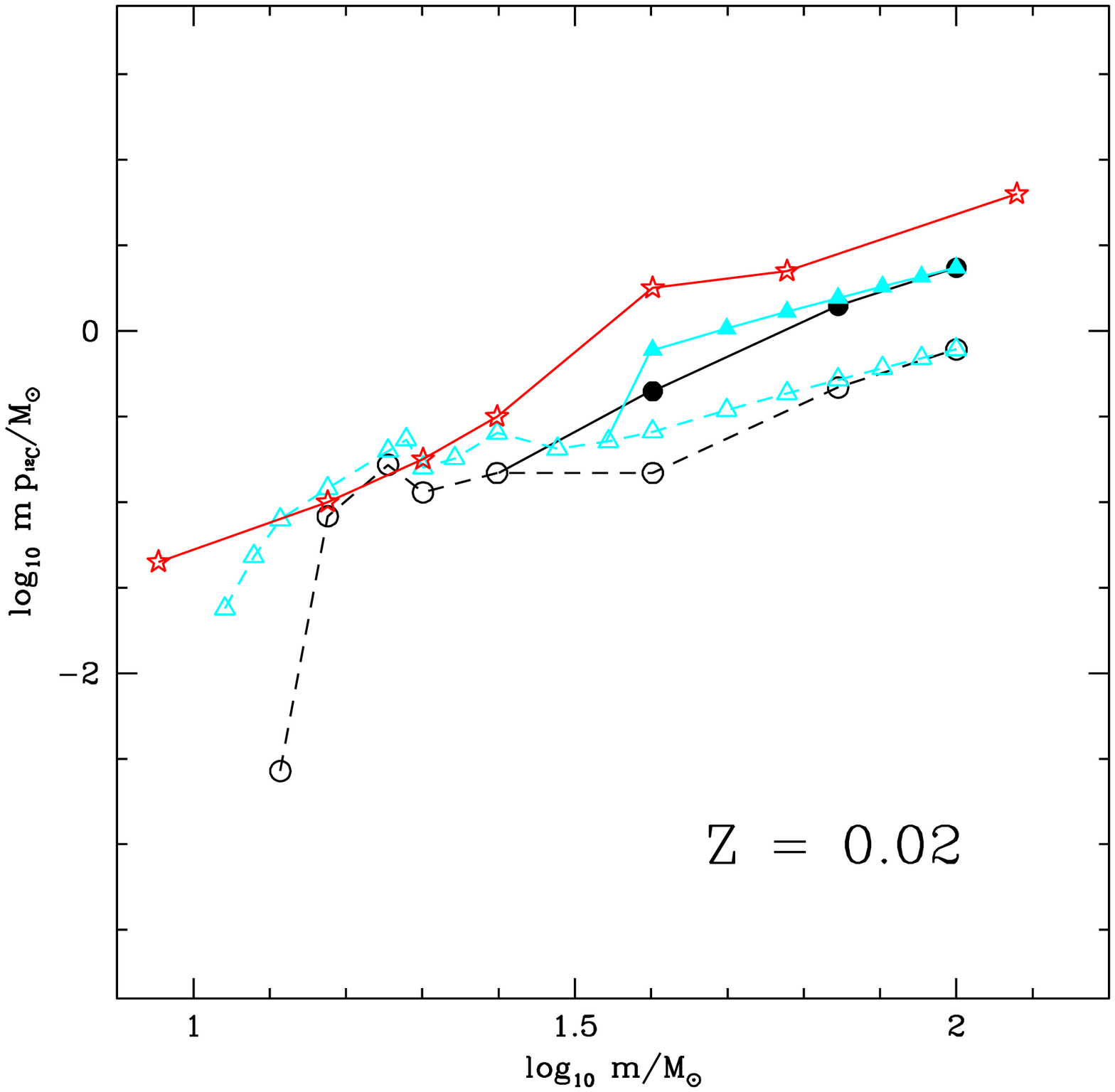,width=14cm,height=14cm} }
\caption{This figure shows a comparison of different yields of carbon from massive stars, computed
for solar metallicity (open circles: TNH; 
filled circles: TNH with carbon yields multiplied by 3;
open triangles: WW; filled triangles: WW with carbon yields
multiplied by 3; stars: Meynet \& Maeder 2002b).}
\end{figure}

\newpage

\begin{figure}
\figurenum{2}
\centerline{\psfig{figure=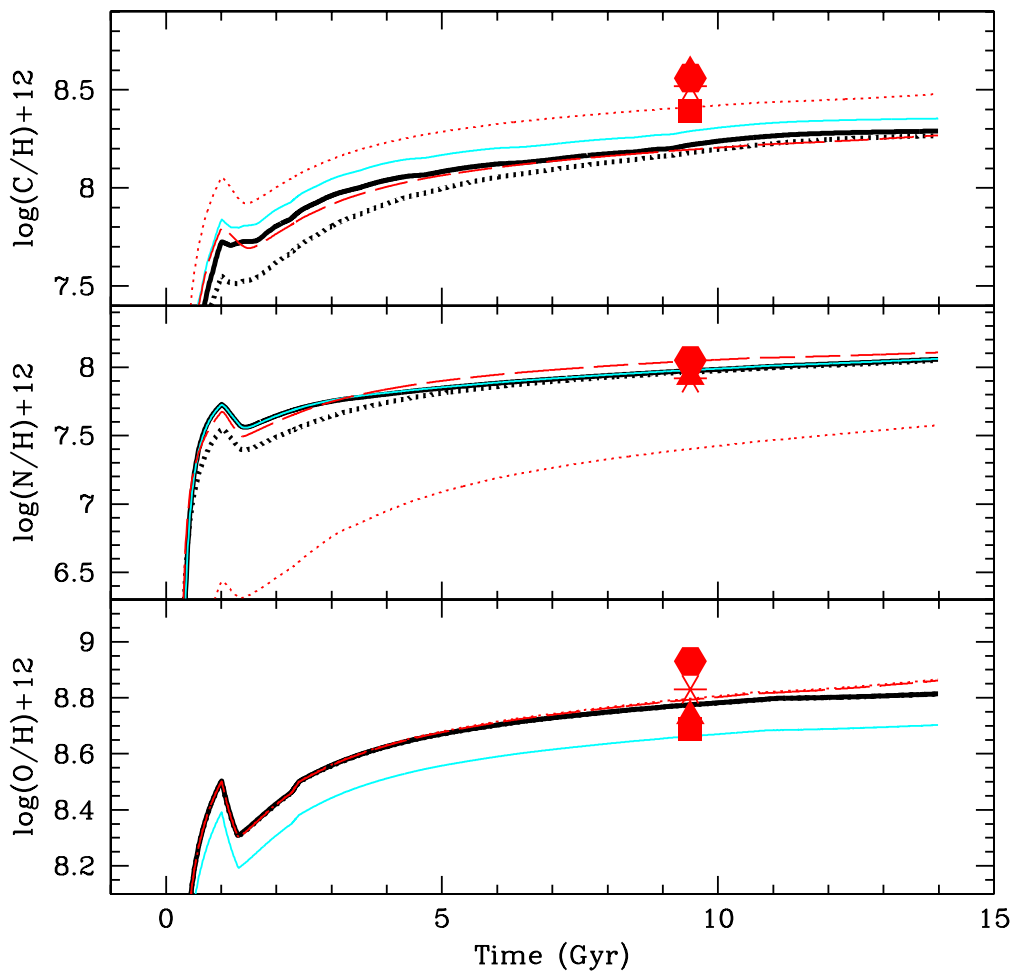,width=14cm,height=14cm} }
\caption{Time evolution of C, N and O in the solar vicinity. The big symbols show the different
location of the Sun in this diagram when adopting Anders \& Grevesse (1989 - hexagon), 
Grevesse \& Sauval (1998 - star) or Holweger (2001 - triangle) and Allende Prieto et al. (2001, 2002, square). The lines show models 1 (thin dotted line), 2 (thin long-dashed line), 3 (thick dotted-line), 4 (thick solid line) and 5 (thin solid line). Note the good agreement of model 5 (computed with WW yields) with the new solar value of Allende Prieto.}
\end{figure}

\newpage

\begin{figure}
\figurenum{3}
\centerline{\psfig{figure=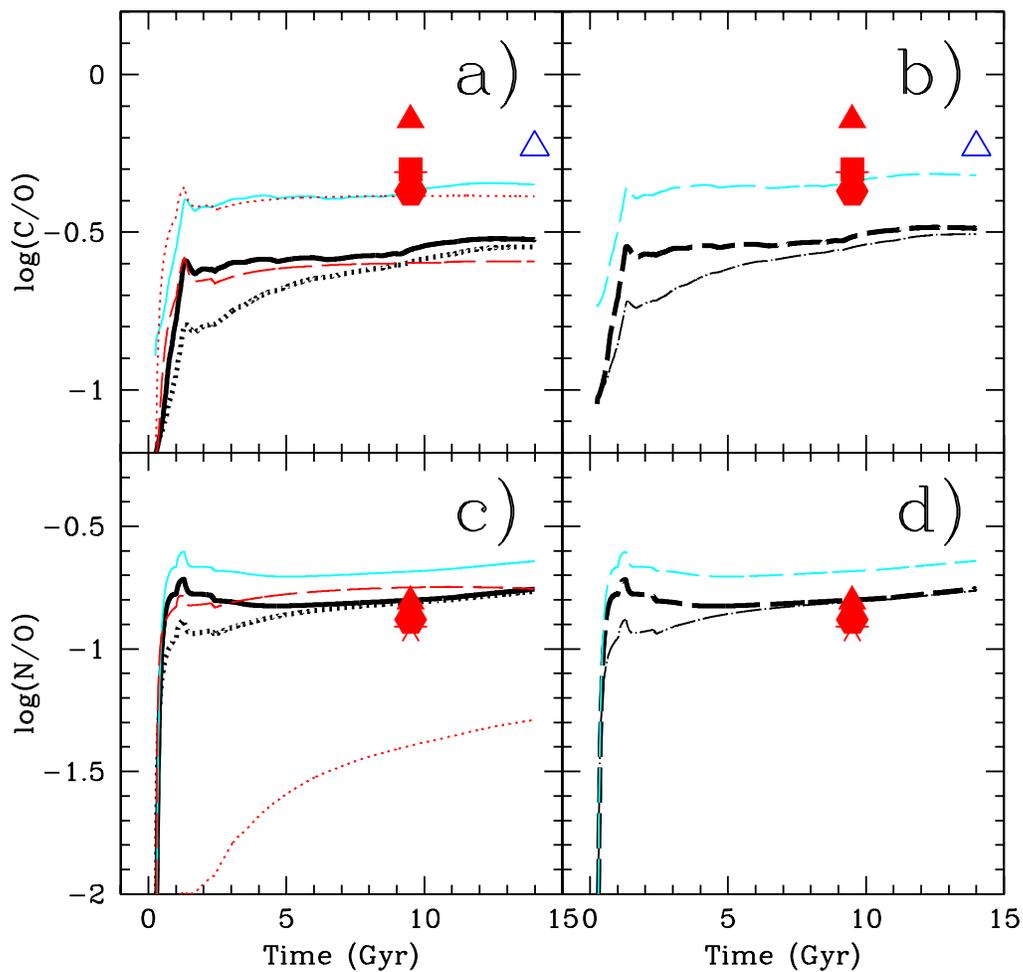,width=14cm,height=14cm} }
\caption{Time evolution of C/O and N/O in the solar vicinity. The big symbols are labeled as in Figure 2. The Orion measurement by Peimbert (1999) is also shown (open triangle). Note that C measurements in the sun include both isotopes ($^{12}$C and $^{13}$C), while our curves refer only to $^{12}$C. In panels a) and c) we show models 1 to 5 (labeled as in Figure 2), while in panels b) and d) we show
models 3a (dash-dotted line), 6 (long-dashed thick line) and 7 (long-dashed thin line).}
\end{figure}

\newpage

\begin{figure}
\figurenum{4}
\centerline{\psfig{figure=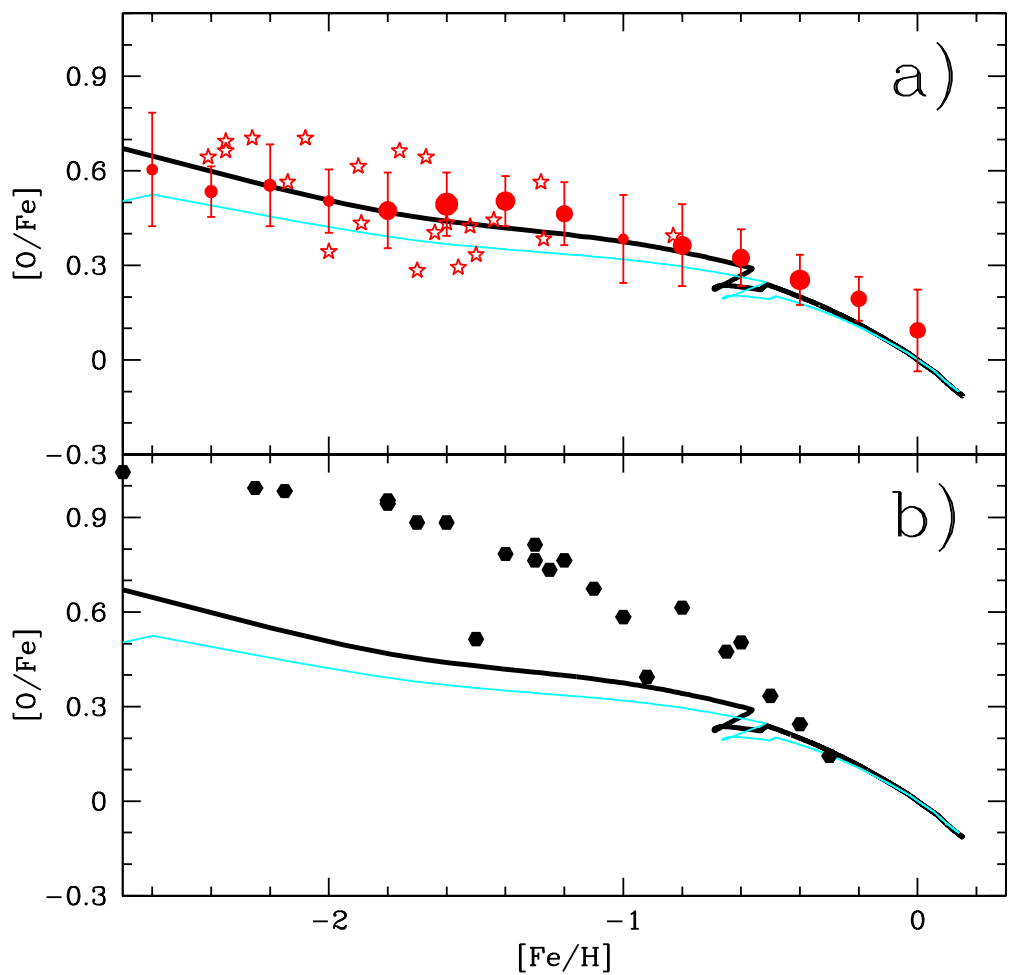,width=14cm,height=14cm} }
\caption{Figure 4a shows our predictions for models 4 and 5 (thick and thin lines,
respectively) compared with the recent observations (stars) by Mel\'endez \& Barbuy (2002).
The dots show the compilation of the
best data in the literature as done by Mel\'endez \& Barbuy (2002) and this
includes the very-metal-poor object recently measured
by Cayrel et al. (2001) who measured
[OI] for a very-metal-poor star by using VLT. As discussed in the text, 
[OI] is the most reliable way to infer the oxygen abundance
and so more weight should be given to this particular measurement. Figure 4b shows the 
abundance data obtained from UV OH lines by Israelian et al. (1998) and Booesgard et al. (1999).}
\end{figure}

\newpage

\begin{figure}
\figurenum{5}
\centerline{\psfig{figure=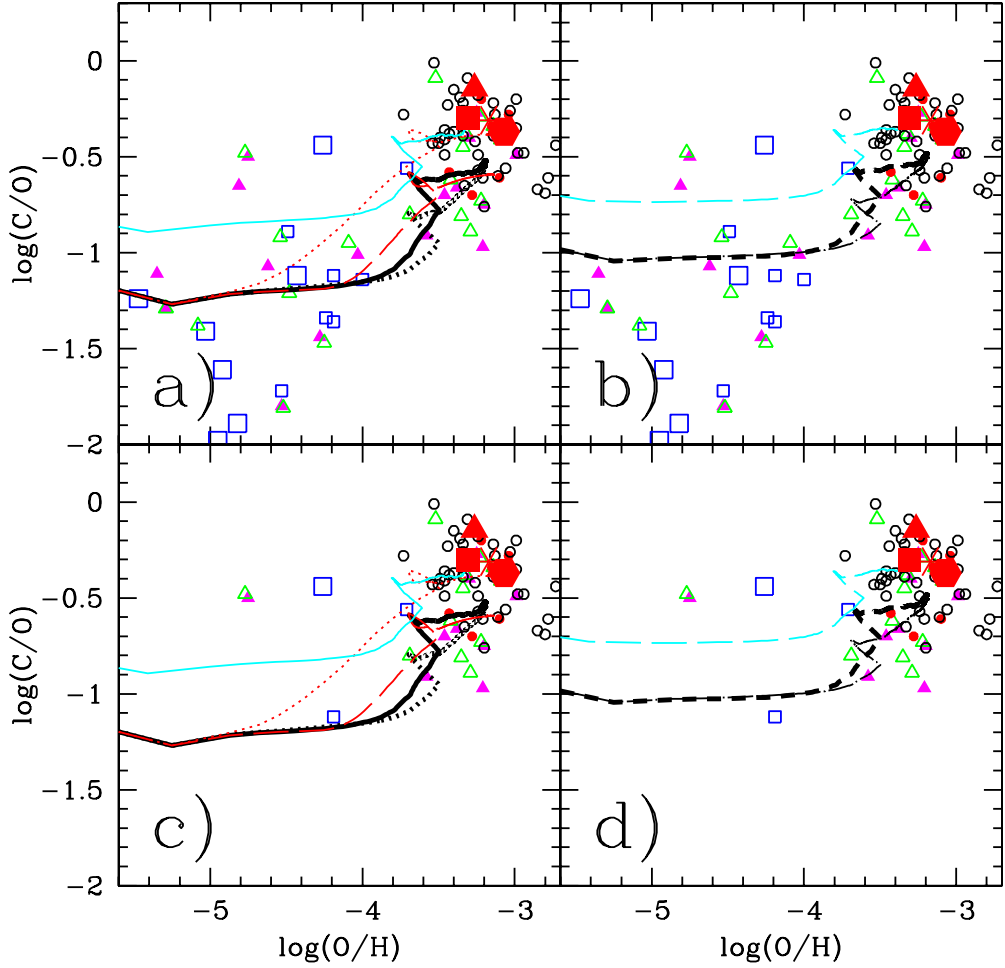,width=14cm,height=14cm} }
\caption{In panels a) and b) 
we show the carbon and oxygen abundance measurements
obtained by Mel\'endez et al. 2001 (small open squares),
Mel\'endez \& Barbuy 2002 (big open squares), 
reanalysis by Carretta et al (2000) of the
data of Edvardsson et al. (1993) (open circles), Carretta
et al. (2000) for oxygen values obtained as the mean of the abundances
obtained with [OI] and OI oxygen lines (open triangles) and 
for oxygen obtained only from [OI] measurements (filled triangles). In panels
c) and d) only the data for dwarf stars are shown. In panels a) and c)
we show the predictions of models 1 to 5 (labeled as in Fig.2), whereas in
panels b) and d) we show models 3a (dash-dotted line), 6 (long-dashed thick 
line) and 7 (long-dashed thin line).}
\end{figure}

\newpage

\begin{figure}
\figurenum{6}
\centerline{\psfig{figure=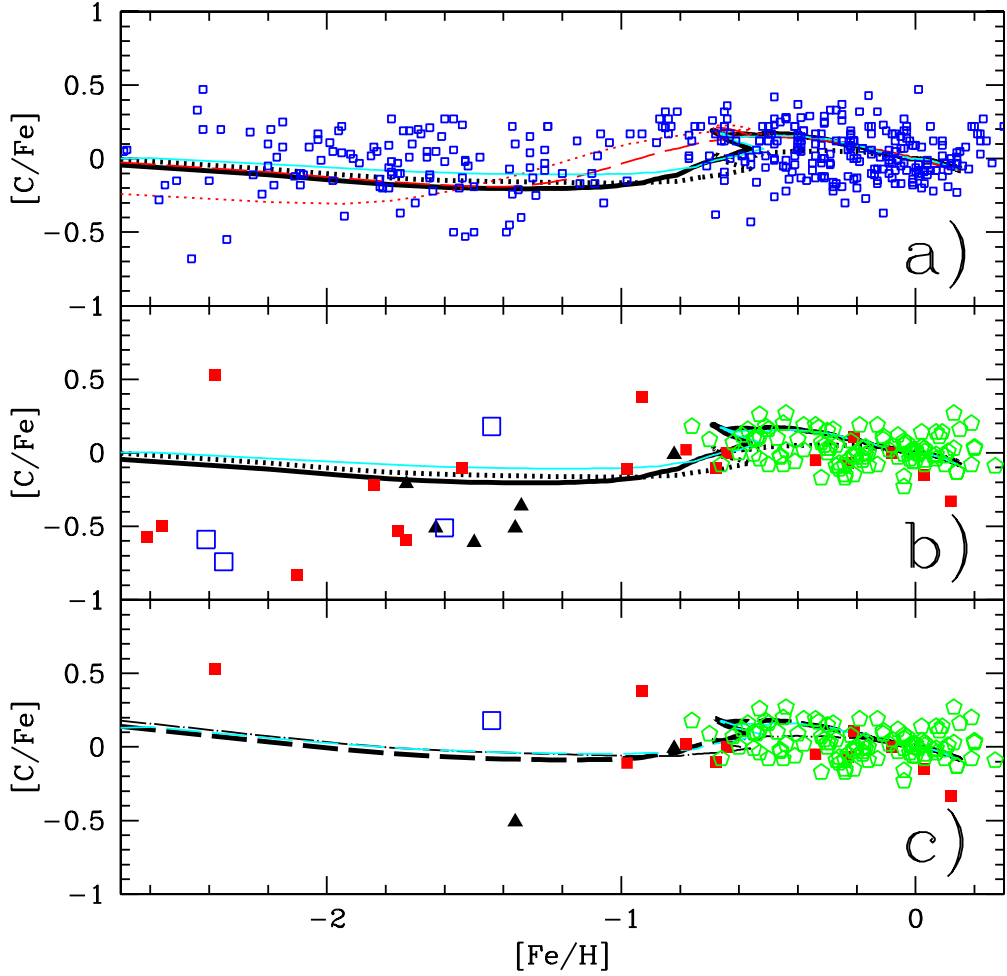,width=14cm,height=14cm} }
\caption{[C/Fe] vs. [Fe/H] plot: a) we 
show the model predictions for models 1 to 5 listed in Table 2a (labeled as in Figure 2)
compared with the compilation of Chiappini et al. (1999 - open squares); 
b) we show only the best 
measurements for carbon: Mel\'endez et al. (2001 - big open squares; 2002 - filled triangles), 
Carretta et al. (2000,
small filled squares) and the reanalysis by Carretta et al. (2000 - open pentagons) of the data of 
Edvardsson et al. (1993). Models 3, 4 and 5 are also shown; 
c) dwarf stars are plotted along with the predictions of models 3a, 6 and 7 (labeled as in Figure 5).}
\end{figure}

\newpage

\begin{figure}
\figurenum{7}
\centerline{\psfig{figure=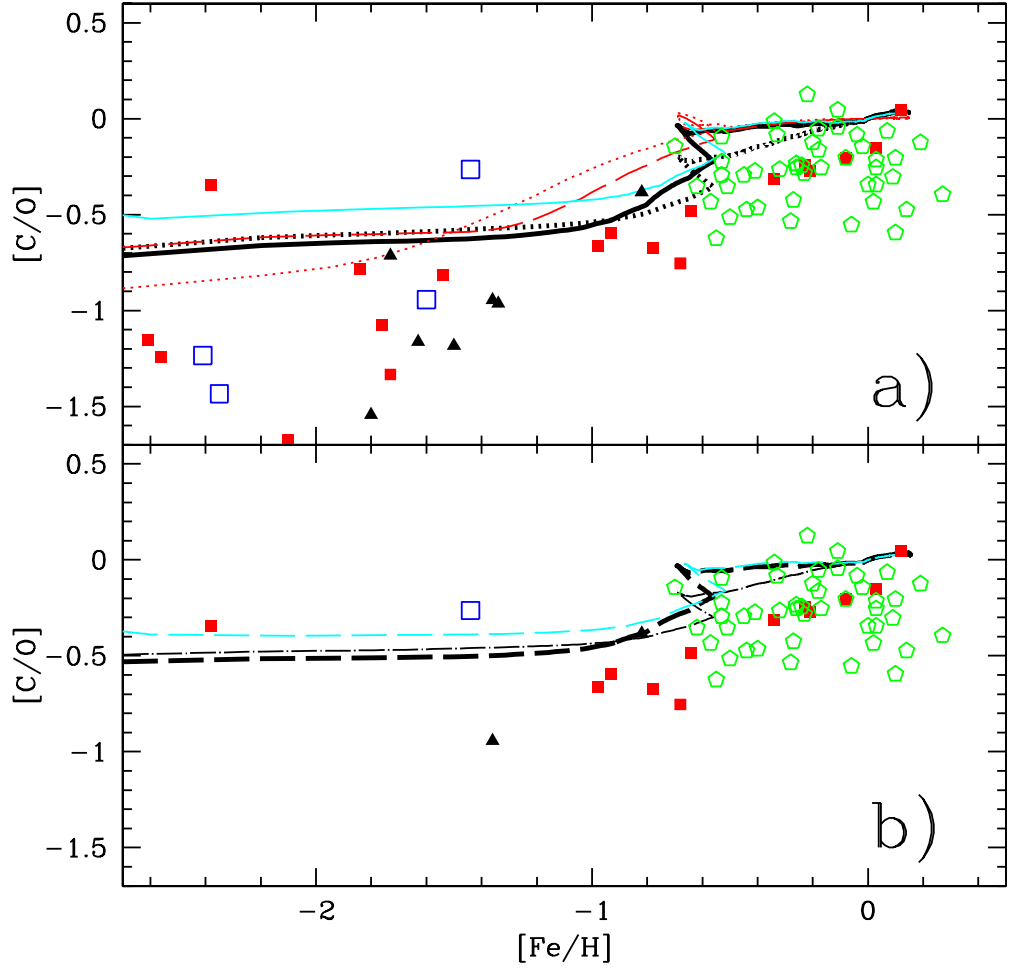,width=14cm,height=14cm} }
\caption{[C/O] vs. [Fe/H]. Data are labeled as in Figure 6. In panel b) only dwarf stars are plotted.
Models are labeled as in Figure 6 (models 1 to 5 in panel a and models 3a, 6 and 7 in panel b).}
\end{figure}

\newpage
\begin{figure}
\figurenum{8}
\centerline{\psfig{figure=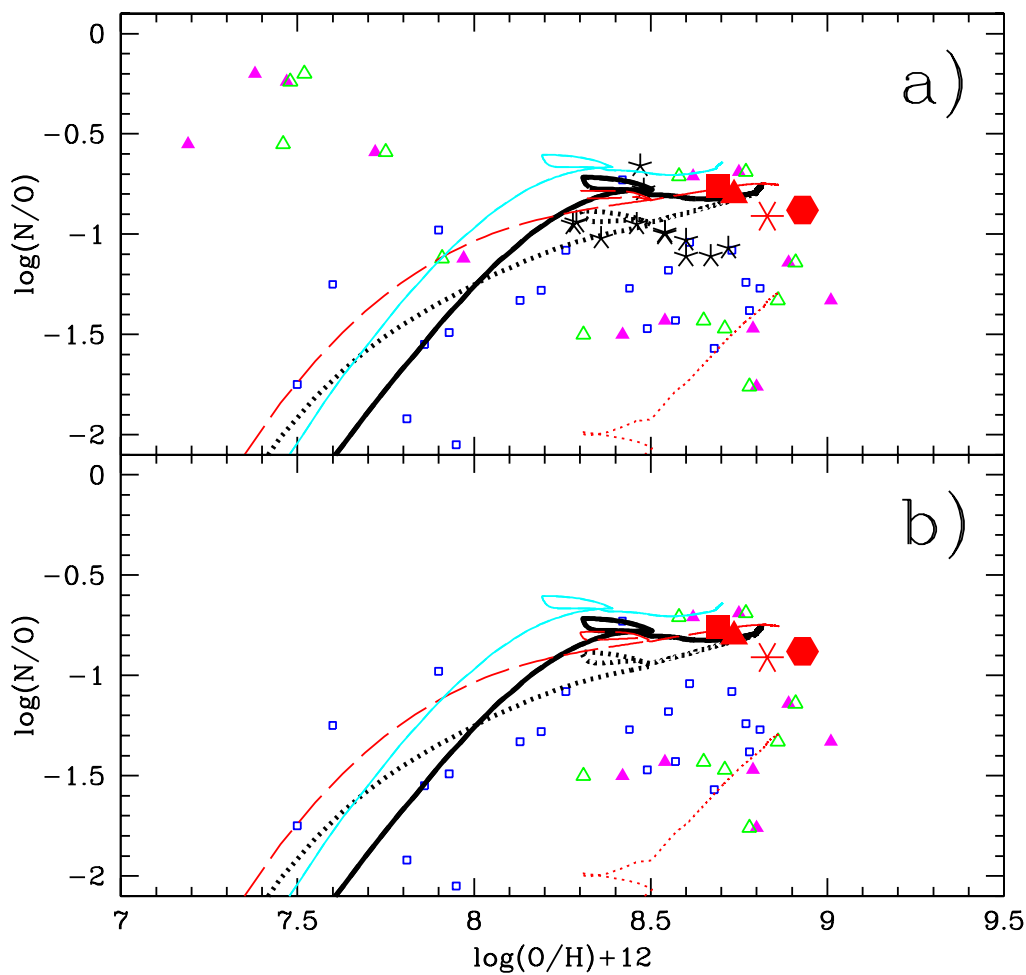,width=14cm,height=14cm} }
\caption{This figure shows log(N/O) vs. log(O/H)+12. In this case very few 
data are available. In the upper panel we plot all the stars with measured
abundances of N and O, while in the lower panel we plot only dwarf stars
(see text). The triangles are as in Figure 5. Small squares are data from 
the compilation of Chiappini et al. (1999).
In this figure we also added some recent measurements of B stars (Daflon et al. 2001: stars).
The big symbols represent different solar measurements, as in previous figures. Models 1 to 5
are shown (labeled as in Figure 2).}
\end{figure}

\newpage

\begin{figure}
\figurenum{9}
\centerline{\psfig{figure=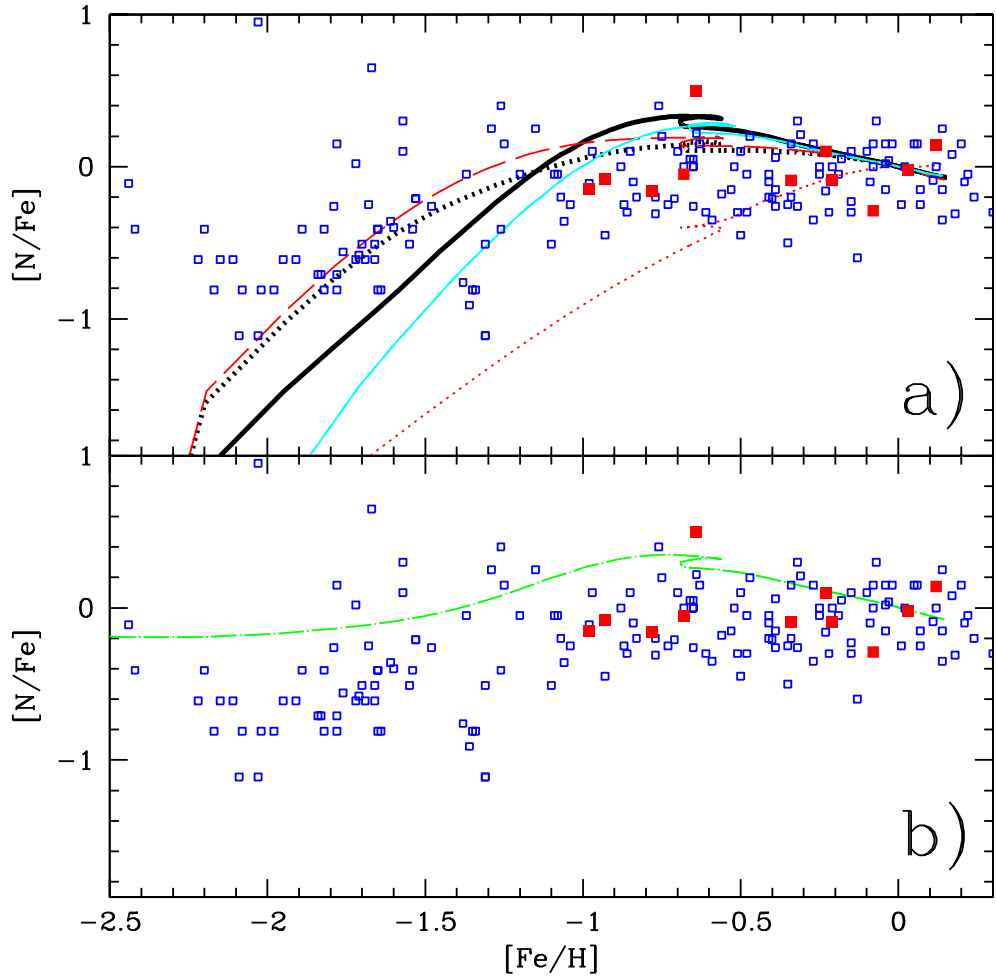,width=14cm,height=14cm} }
\caption{This figure shows [N/Fe] vs. [Fe/H]. The data points are from
Chiappini et al. (1999): open squares; Carretta
et al. (2000): filled squares. From panel a), where models 1 to 5 are plotted, 
it can be seen that the abundance data in the MW are not inconsistent
with a pure secondary N production in massive stars. In panel b) we show the predictions
of model 8 which assumes an important primary N contribution from massive stars.}
\end{figure}

\newpage

\begin{figure}
\figurenum{10a}
\centerline{\psfig{figure=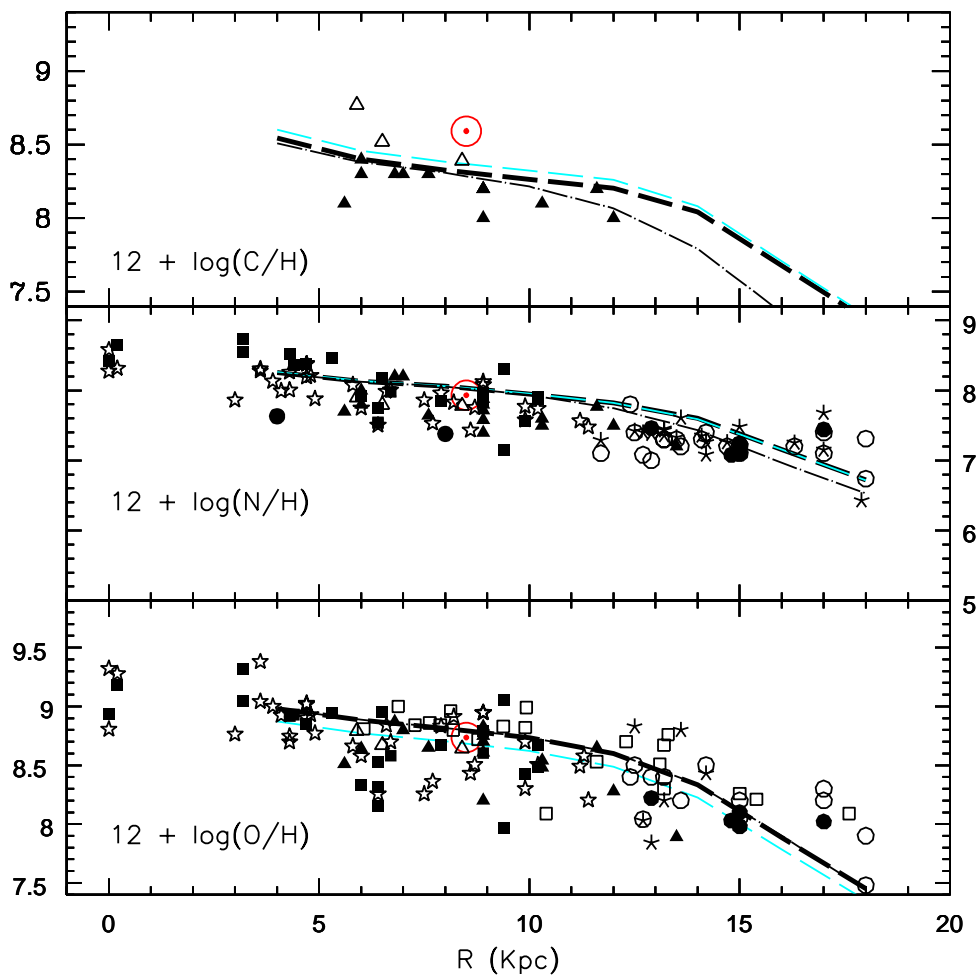,width=14cm,height=14cm} }
\caption{This figure shows the abundance gradients of C, N and O in the Milky Way. In the 
upper panel the data are from Esteban et al. (1999 - open triangles, HII regions) and 
Gummersbach et al. (1998 - filled triangles, B stars).
In the middle and lower panels
the data are from Gummersbach et al. (1998 - filled triangles, B stars), Simpson et al. 
(1995 - filled squares, HII regions), Vilchez \& Esteban (1996 - open circles, HII regions), Afflerbach
et al. (1997 - open stars, HII regions), Rudolph et al. (1997, filled circles, HII regions), Fich \& 
Silkey (1991 - asterisks, HII regions). 
The curves show the predictions of models 3a (dash-dotted line),
6 (long-dashed thick line) and 7 (long-dashed thin line). The Sun position is also
shown by the usual symbol.}
\end{figure}

\newpage

\begin{figure}
\figurenum{10b}
\centerline{\psfig{figure=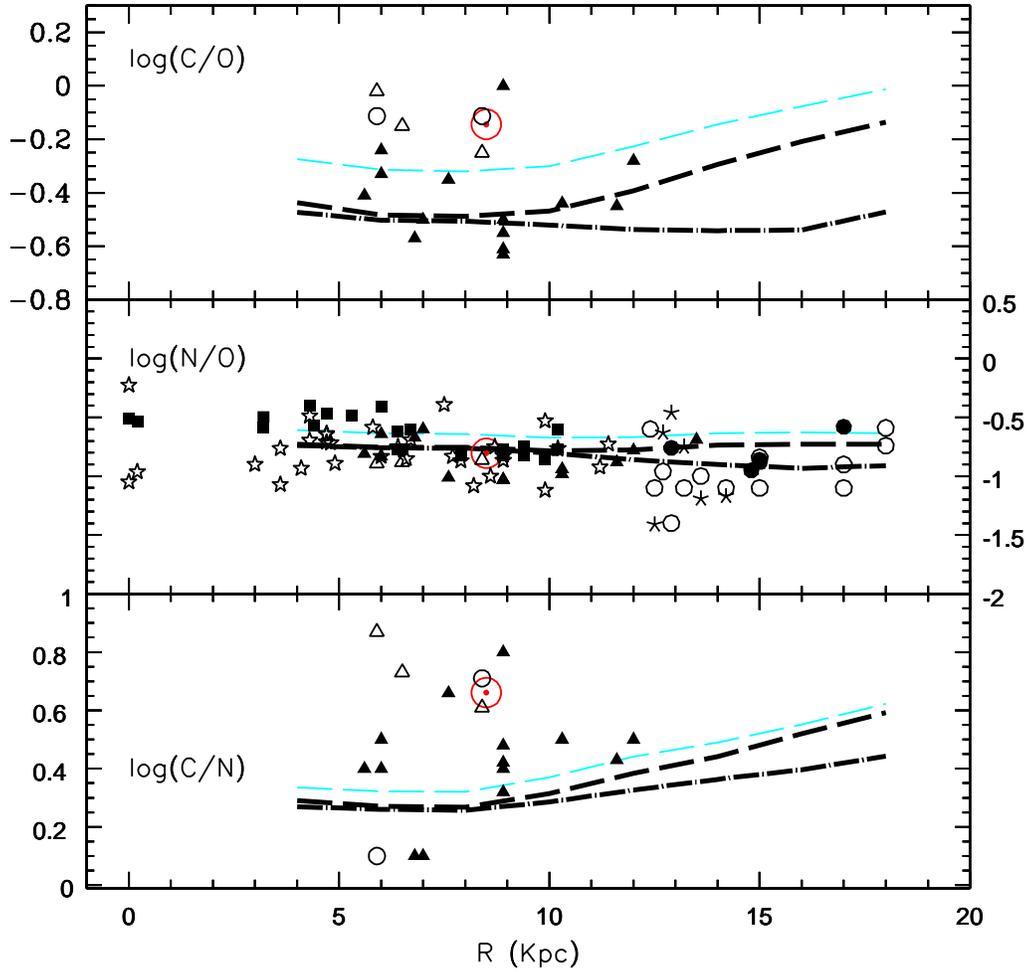,width=14cm,height=14cm} }
\caption{Abundance gradients of C/O, N/O and C/N in the Milky Way. Data points and models 
are labeled as in Figure 10a. In the upper and lower panels we add the data by
Tsamis et al. (2002 - open circles, HII regions).}
\end{figure}

\newpage

\begin{figure}
\figurenum{11}
\centerline{\psfig{figure=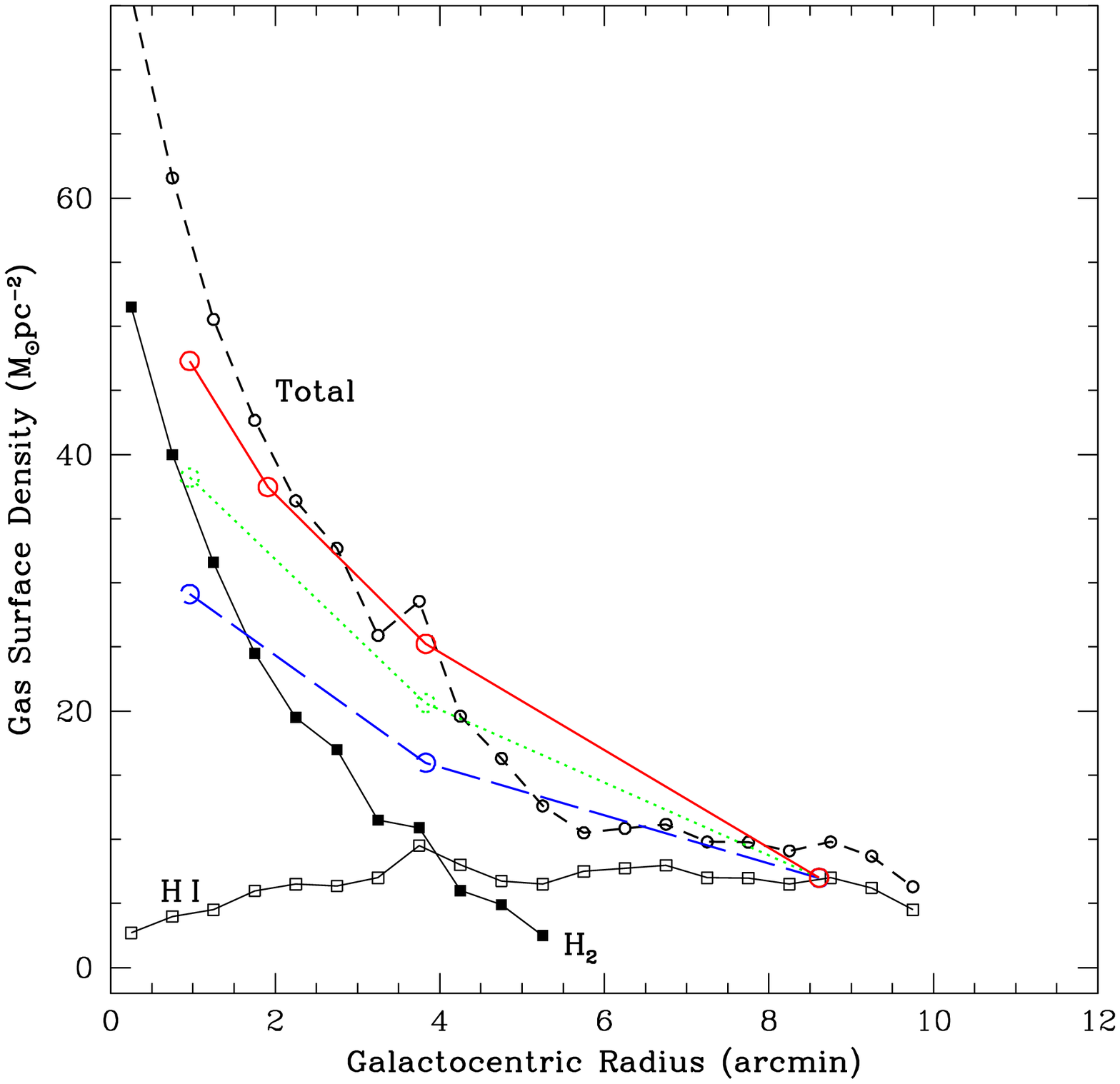,width=14cm,height=14cm} }
\caption{Total gas surface density profile in M101: observed - small
open symbols connected by short-dashed line; models: large circles
connected by solid, long-dashed and dotted lines for models 
with different assumptions for the central mass density (see text). 
The total gas surface density profile is given by 1.4 (HI + H2), where the atomic and molecular
gas profiles come from Kenney et al. (1991) and are shown by the open and 
full squares, respectively. }
\end{figure}

\begin{figure}
\figurenum{12}
\centerline{\psfig{figure=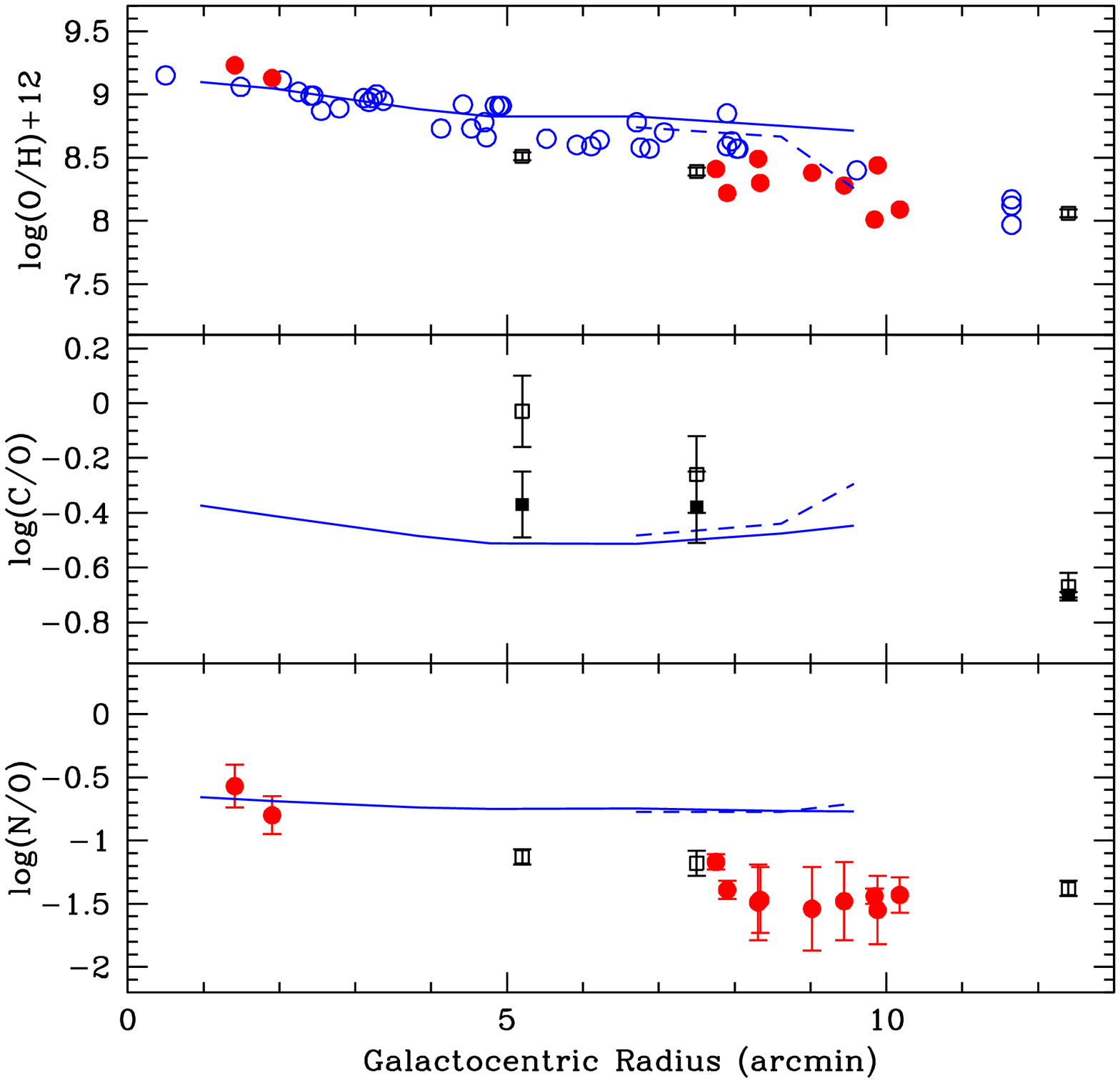,width=14cm,height=14cm} }
\caption{Abundance gradients of O/H, C/O and N/O in M101. The lines show
our model predictions. The dashed line shows a model with a larger value
for the threshold gas mass density with respect to the model represented by the solid line.
The data are from Kennicutt and Garnett (1996, open circles);
van Zee et al. (1998a, full circles), Garnett et al. (1999, open squares). For 
C/O Garnett et al. (1999) show two values depending on the adopted reddening correction (open
squares for $R_v=$3.1 and filled squares for $R_v=$5.0 - see Garnett et al. 1999 for details)  }
\end{figure}

\begin{figure}
\figurenum{13a}
\centerline{\psfig{figure=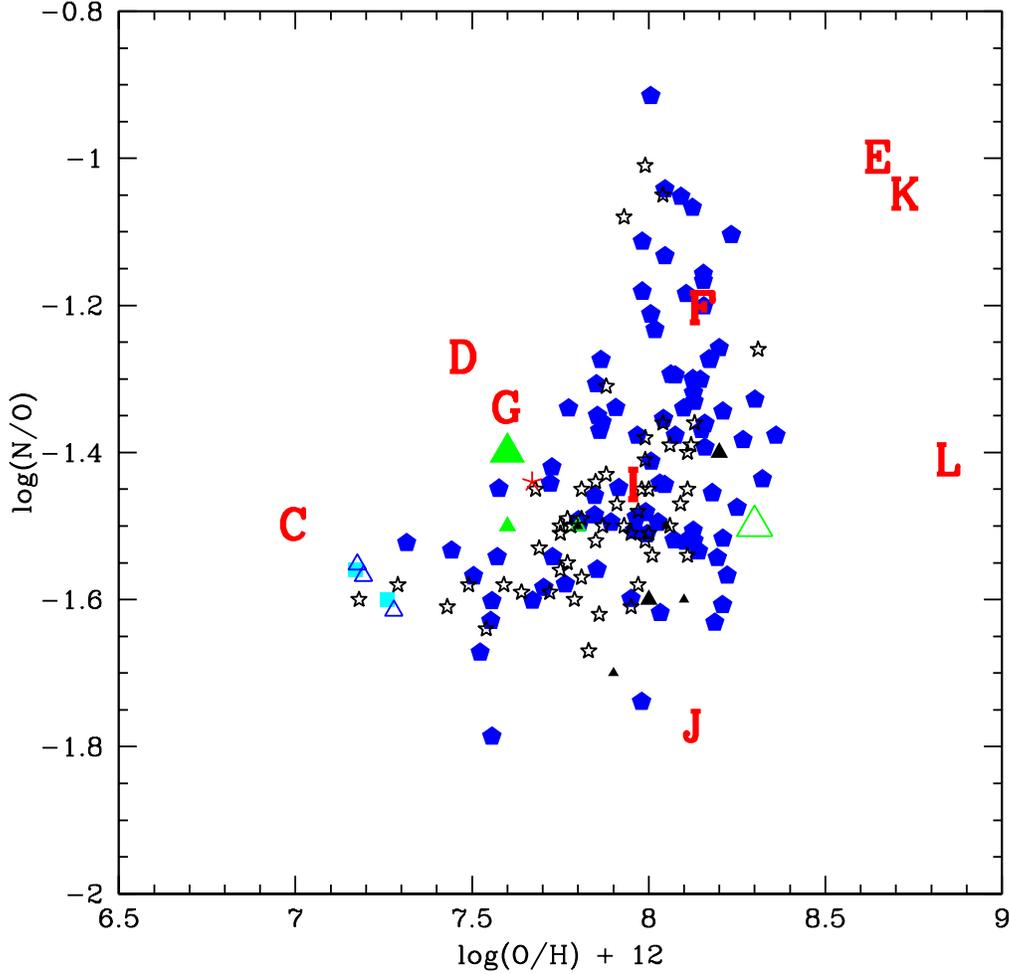,width=14cm,height=14cm} }
\caption{Abundances of N and O in dwarf irregular galaxies. The letters represent
different chemical evolution models - see text and Table 5.
The data are from Garnett et al. 1997b (filled squares - IZw 18); Kobulnicky and Skillman 1996 (filled
pentagons - HII galaxies, open triangles - IZw18); van Zee et al. 1997 (filled triangles - low surface
brightness dwarf galaxies. The large filled triangle is an object showing a low
SFR at the present time - 0.0015 M$_{\odot}$yr$^{-1}$, while the open large triangle is an object which
has a larger SFR - 0.35 M$_{\odot}$yr$^{-1}$); Izotov \& Thuan 1999 (stars); Kennicutt \& Sillkman 2001 (asterisk - dwarf irregular galaxy DDO 154).}
\end{figure}

\begin{figure}
\figurenum{13b}
\centerline{\psfig{figure=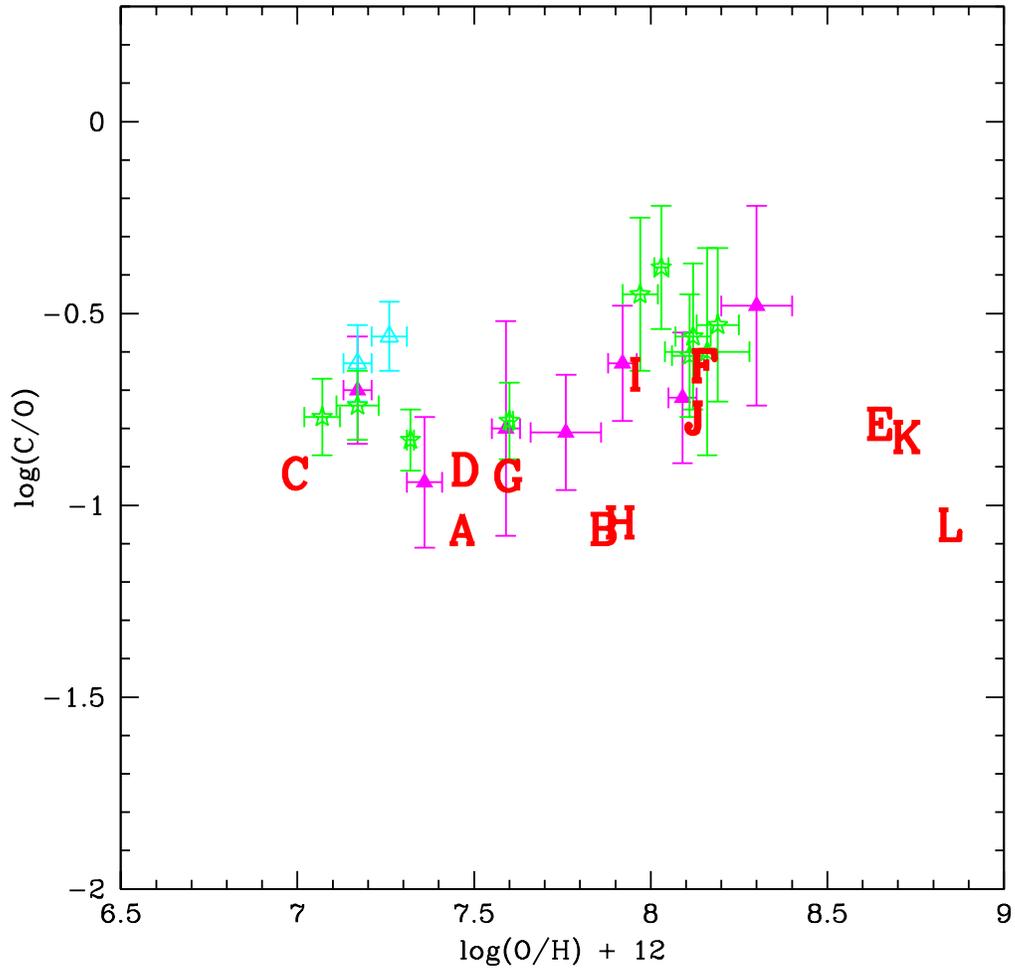,width=14cm,height=14cm} }
\caption{Abundances of C and O available for dwarf irregular galaxies. 
This sample includes blue compact galaxies,
low surface brightness dwarf galaxies and the most metal-poor known galaxy, IZw18.
The data are from Garnett et al. (1995a,b - filled triangles); Garnett et al. (1997 - IZw18 - open triangles) and Izotov \& Thuan (1999 - stars).
Letters are as in figure 13a.}
\end{figure}

\begin{figure}
\figurenum{14}
\centerline{\psfig{figure=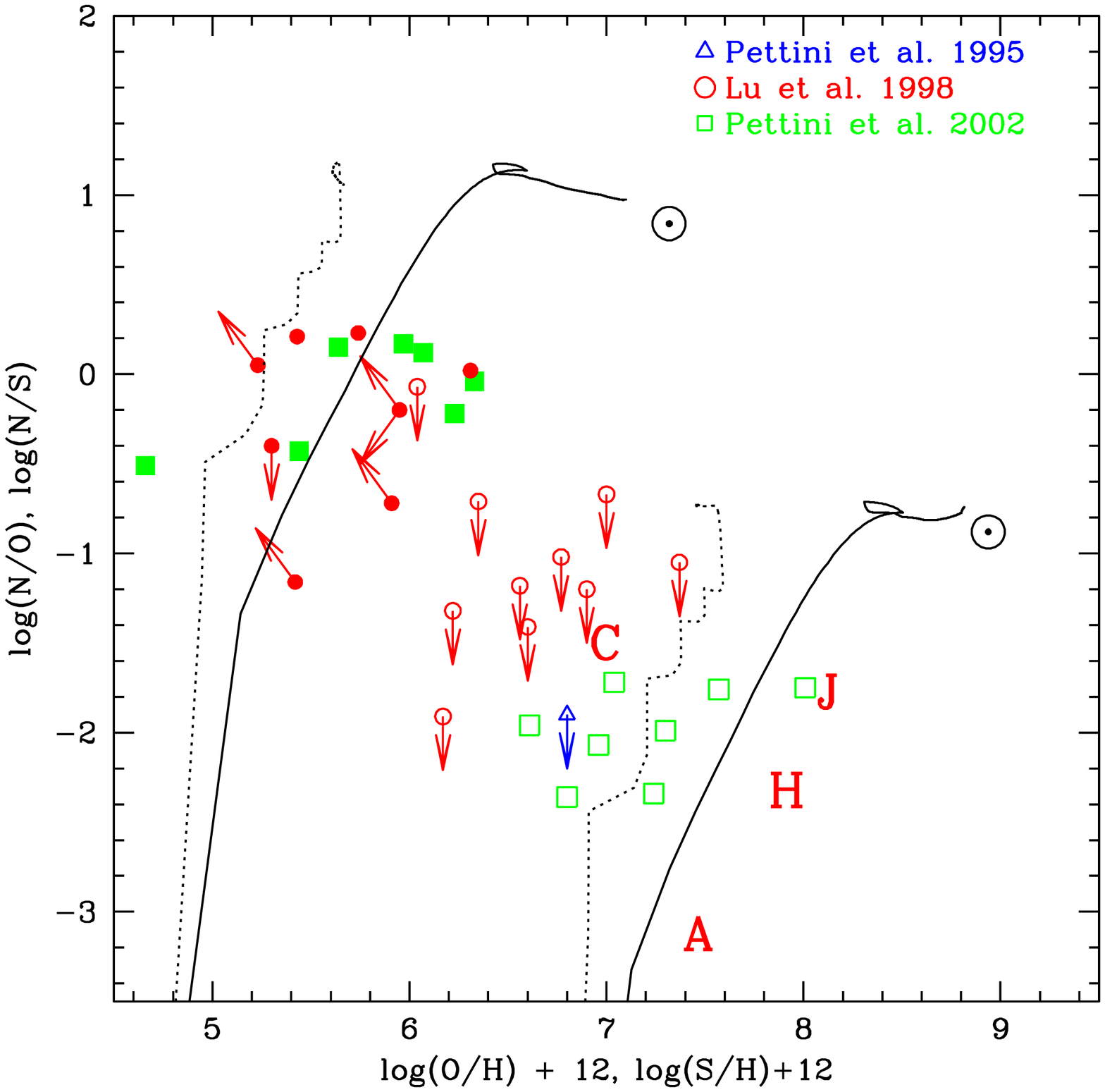,width=14cm,height=14cm} }
\caption{Filled symbols correspond to objects
where S was measured, while open symbols refer to O measurements. The arrows
indicate when a measurement is an upper/lower limit. 
The solar values for both N/S (upper left corner) and N/O
(low right corner) are also shown by the usual symbol. Those data are compared with chemical evolution models 
for the MW at the solar vicinity (solid line) and at 18 kpc from the Galactic center (dotted line).
The letters A, C, H and J correspond to different models for dwarf galaxies - see Sect.4.2.2.}

\end{figure}


\begin{thebibliography}{}

\bibitem{}
Afflerbach, A., Churchwell, E., Werner, M. W. 1997, ApJ 478, 190
\bibitem{}
Aloisi, A., Tosi, M., Greggio, L. 1999, ApJ, 118, 302
\bibitem[]{}
Anders, E., \& Grevesse, N. 1989, Geochim. Cosmochim. Acta, 53, 197
\bibitem{}
Allende Prieto, C., Lambert, D. L., Asplund, M. 2002, ApJ, 573, L137
\bibitem{}
Allende Prieto, C., Lambert, D. L., Asplund, M. 2001, ApJ, 556, L63
\bibitem{}
Asplund, M. \& Garcia P\'erez, A.E. 2001, A\&A, 372, 601
\bibitem{}
Boesgaard, A. M., King, J. R., Deliyannis, C. P., \& Vogt, S. S. 1999, AJ, 117, 492
\bibitem{}
Boissier, S. \& Prantzos, N. 2000, MNRAS, 312, 398
\bibitem[]{}
Boothroyd, A.I., Sackmann, I.-J., \& Ahern, S.C. 1993, ApJ, 416, 762
\bibitem[]{}
Boothroyd, A.I., Sackmann, I.-J., \& Wasserburg, G.J 1995, ApJ, 442, L21
\bibitem{}
Bradamante, F., Matteucci, F., D'Ercole, A. 1998, A\&A 337, 338
\bibitem{}
Calura, F., Matteucci, F., Vladilo, G. 2002, MNRAS (submitted)
\bibitem{}
Carbon, D. F., Barbuy, B., Kraft, R. P., Friel, E. D., Suntzeff, N. B. 1987, PASP, 99, 335
\bibitem{}
Carigi, L. 2000, Rev. Mex. Astron. Astrof., 36, 171
\bibitem{}
Carretta, E., Gratton, R. G., Sneden, C. 2000, A\&A, 356, 238
\bibitem{}
Cayrel, R., Andersen, J., Barbuy, B., Beers, T. C., Bonifacio, P., Fran\c cois, P., Hill, V., Molaro, P., Nordstr\"om, B., Pletz, B., Primas, F., Spite, F., Spite, M. 2001, New Astronomy Rev., 45, 533
\bibitem{}
Chiappini, C., Matteucci, F., Romano, D. 2001, ApJ, 554, 1044 (CMR2001)
\bibitem{}
Chiappini, C., Matteucci, F., Padoan, P. 2000, ApJ, 528, 711
\bibitem{}
Chiappini, C., Matteucci, F., Beers, T. C., Nomoto, K. 1999, ApJ, 515, 226
\bibitem{}
Chiappini, C., Matteucci, F., \& Gratton, R. 1997, ApJ, 477, 765 (CMG97)
\bibitem{}
Cunha, K. \& Lambert, D. L. 1992, ApJ, 399, 586
\bibitem{}
Daflon, S., Cunha, K., Butler, K., Smith, V. V. 2001, ApJ, 563, 325
\bibitem{}
Depagne, E., Hill, V., Spite, M. et al. 2002, A\&A, 390, 187
\bibitem{}
Diaz, A. I., Tosi, M. 1986, A\&A, 158, 60
\bibitem{}
Edmunds, M. G. \& Pagel, B. E. J. 1994, MNRAS, 211, 507
\bibitem{}
Edvardsson, B., Andersen, J., Gustafsson, B., Lambert, D.L., Nissen, P. E. \& Tomkin, J.
1993, A\&A, 275, 101
\bibitem{}
Elmegreen, B. G. 1999, ApJ, 517, 103
\bibitem{}
Esteban, C., Peimbert, M., Torres-Peimbert, S., Garcia-Rojas, J., Rodriguez, M. 1999, ApJS, 120, 113
\bibitem{}
Ferguson, A. 2002, in ``The Evolution of Galaxies II. Basic Building Blocks'', eds  
M. Sauvage, G. Stasinska and D. Schaerer, Kluwer, p. 119 
\bibitem[]{}
Ferguson, A.M.N., Gallagher, J.S., \& Wyse, R.F.G. 1998, AJ, 116, 673
\bibitem{}
Fich, M., Silkey, M. 1991, ApJ, 366, 107
\bibitem{}
Fuhrmann, K. 1998, A\&A, 338, 161
\bibitem{}
Fulbright, J. P. \& Kraft, R. P. 1999, AJ, 118, 527
\bibitem[]{}
Garnett, D. R., Shields, G. A., Peimbert, M., Torres-Peimbert, S., Skillman, E. D., Dufour, R. J., Terlevich, E., \& Terlevich, R. J. 1999, ApJ, 513, 168
\bibitem[]{}
Garnett, D. R., Shields, G. A., Skillman, E. D., Sagan, S. P., \& Dufour, R. J. 1997a, ApJ, 489, 63
\bibitem{}
Garnett, D. R., Skillman, E. D., Dufour, R. J., Shields, G. A. 1997b, ApJ, 481, 174
\bibitem{}
Garnett, D. R., Skillman, E. D., Dufour, R. J., Peimbert, M., Torres-Peimbert, S., Terlevich, R., Terlevich, E., Shields, G. A. 1995a, ApJ, 443, 64
\bibitem{}
Garnett, D. R., Dufour, R. J., Peimbert, M., Torres-Peimbert, S., Shields, G. A., Skillman, E. D., Terlevich, E., Terlevich, R. J. 1995b, ApJ, 449, L77
\bibitem{}
Gerritsen, J. P. E., Icke, V. 1997, A\&A, 325, 972
\bibitem{}
Gratton, R.G., Carretta, E., Matteucci, F., \& Sneden, C. 2000, A\&A, 358, 671
\bibitem{}
Grevesse, N., \& Sauval, A.J. 1998, Space Sci. Rev., 85, 161
\bibitem{}
Grevesse, N., Noels, A., Sauval, A. J. 1996, ASP Conf. Ser., vol. 99, p.117
\bibitem[]{}
Groenewegen, M.A.T., \& de Jong, T. 1993, A\&A, 267, 410
\bibitem[]{}
Groenewegen, M.A.T., van den Hoek, L.B., \& de Jong, T. 1995, A\&A, 293, 381
\bibitem{}
Gummersbach, C. A., Kaufer, A., Sch\"afer, D. R., Szeifert, T., Wolf, B. 1998, A\&A, 338, 881
\bibitem{}
Gustafsson, B., Karlsson, T., Olsson, E., Edvardsson, B. \& Ryde, N. 1999, A\&A, 342, 426
\bibitem[]{}
Henry, R.B.C., Edmunds, M.G., K\"oppen, J. 2000, ApJ, 541, 660
\bibitem{}
Holweger, H. 2001, in Joint SOHO/ACE Workshop: Solar and Galactic Composition, ed. R. F. 
Willmmer-Schweingruber, Am. Inst. of Physics Conf. Proc., vol 598, p. 23
\bibitem{}
Israelian, G., Rebolo, R., Garcia L\'opez, R.J., Bonifacio, P., Molaro, P., Basri, G. \&
Shchukina, N. 2001, ApJ, 551, 833
\bibitem{}
Israelian, G., Garcia L\'opez, R. J., Rebolo, R. 1998, ApJ, 507, 805
\bibitem{}
Izotov, Y. I., Chaffee, F. H., Foltz, C. B., Green, R. F., Guseva, N. G., Thuan, T. X. 1999, ApJ, 527, 757
\bibitem{}
Izotov, Y. I., Thuan, T. X. 1999, ApJ, 511, 639
\bibitem[]{}
Kenney, J.D.P., Scoville, N.Z., \& Wilson, C.D. 1991, ApJ, 366, 432
\bibitem{}
Kennicutt, R. C, Skillman, E. D. 2001, AJ, 121, 1461
\bibitem{}
Kennicutt, R. C. 2001, in Galaxy Disks and Disk Galaxies, eds. J. G. Funes, S. J. and E. M. Corsini, ASP Conf. Ser., Vol 230, pp. 291-298
\bibitem{}
Kennicutt, R. C., Skillman, E. D. 2001, AJ, 121, 1461
\bibitem{}
Kennicutt, R. C. 1998, ApJ, 498, 541
\bibitem{}
Kennicutt, R. C., Garnett, D. R. 1996, ApJ, 436, 504
\bibitem{}
Kennicutt, R. C., Skillman, E. D. 1996, ApJ, 462, 147
\bibitem{}
Kotoneva, E., Flynn, C., Chiappini, C., Matteucci, F. 2002, MNRAS, in press (astro-ph/0206446)
\bibitem{}
Laird, J. B. 1985, ApJ 289, 556
\bibitem{}
Lambert, D. 2001, in ``Oxygen Abundances in Old Stars and Implications to Nucleosynthesis and Cosmology'', 24th meeting of the IAU, Joint Discussion 8
\bibitem{}
Larsen, T. I., Sommer-Larsen, J., Pagel, B. E. J. 2001, MNRAS, 323, 555
\bibitem{}
Liang, Y. C., Zhao, G., Shi, J. R. 2001, A\&A, 374, 936
\bibitem{}
Lu, L., Sargent, W. L. W., Barlow, T. A. 1998, AJ 115, 55
\bibitem[]{}
Maeder, A. 1992, A\&A, 264, 105
\bibitem{}
Matteucci, F., Chiappini C. 2001, New Astron. Rev., 45, 567
\bibitem[]{}
Matteucci, F., Recchi, S. 2001, ApJ, 558, 351
\bibitem{}
Matteucci, F. 1986, MNRAS, 221, 911
\bibitem{}
Matteucci, F. \& Greggio, L. 1986, A\&A, 154, 279
\bibitem{}
Matteucci, F. \& Tosi, M. 1985, MNRAS 217, 391
\bibitem{}
Matteucci, F., Chiosi, C. 1983, A\&A, 123, 121
\bibitem{}
Mel\'endez, J., Barbuy, B. 2002, ApJ 575, 474
\bibitem{}
Mel\'endez, J., Barbuy, B., Spite, F. 2001, ApJ, 556, 858
\bibitem{}
Meyer, D. M., Jura, M., Cardelli, J. A. 1998, ApJ, 493, 222
\bibitem{}
Meyer, D. M., Cardelli, J. A., Sofia, U. J. 1997, ApJ, 490, L103
\bibitem[]{}
Meynet, G., \& Maeder, A. 2002a, A\&A, 381, L25
\bibitem{}
Meynet, G., \& Maeder, A. 2002b (astro-ph/0205370)
\bibitem{}
Molaro, P. 2002, in XVII IAP Colloquium ``Gaseous Matter in Galaxies and Intergalactic Space'', p. 307
\bibitem{}
Moos, H. W., Sembach, K. R., Vidal-Madjar, A. et al. 2002, ApJS, 140, 3
\bibitem{}
Nissen, P. E. 2002 in ``CNO in the Universe'', Eds. C. Charbonnel, D. Schaerer and G. Meynet, ASP Conf. Ser., in press
\bibitem{}
Nissen, P.E., Primas, F., Asplund, M., Lambert, D.L. 2002, A\&A, 390, 235
\bibitem{}
Nissen, P.E., Primas, F., Asplund, M. 2001, New Astron. Rev,. 45, 545
\bibitem[]{}
Nomoto, K., Hashimoto, M., Tsujimoto, T., Thielemann, F.-K., Kishimoto, N., 
   Kubo, Y., \& Nakasato, N. 1997, Nucl. Phys. A, 616, 79c
\bibitem{}
Okamura, S., Kanazawa, T., Kodaira, K. 1976, PASJ, 28, 329
\bibitem{}
Pagel, B.E.J. 1997, in ``Nucleosynthesis and Chemical Evolution of Galaxies'', Cambridge Univ. Press
\bibitem{}
Pagel, B. E. J., Edmunds, M. G., Fosbury, R. A. E., Webster, B. L. 1978, MNRAS, 184, 569
\bibitem{}
Peimbert, M. 1999, in ``Chemical Evolution from Zero to High Redshift'', ESO Workshop, ed. J. R. Walsh, M. R. Rosa. Berlin: Springer-Verlag, p. 30.
\bibitem{}
Pettini, M., Ellison, S. L., Bergeron, J., Petitjean, P. 2002, A\&A, in press
\bibitem{}
Pettini, M., Lipman, K., Hunstead, R. W. 1995, ApJ, 451, 100
\bibitem[]{}
Plez, B., Smith, V.V., \& Lambert, D.L. 1993, ApJ, 418, 812
\bibitem[]{}
Prantzos, N., Vangioni-Flam, E., \& Chauveau, S. 1994, A\&A, 285, 132
\bibitem{}
Prochaska, J. X., Wolfe, A. M. 2002, ApJ, 566, 68
\bibitem{}
Pilyugin. L. S. 1999, A\&A, 346, 428
\bibitem{}
Rana, N. C. 1991, ARA\&A, 29, 129
\bibitem{}
Recchi, S. 2001, PhD Thesis, University of Trieste, Italy
\bibitem{}
Recchi, S., Matteucci, F., D'Ercole, A. 2001, MNRAS, 322, 800
\bibitem{}
Recchi, S., Matteucci, F., D'Ercole, A., Tosi, M. 2002, A\&A, 384, 799
\bibitem[]{}
Renzini, A., \& Voli, M. 1981, A\&A, 94, 175 (RV)
\bibitem{}
Rudolph, A. L., Simpson, J. P., Haas, M. R., Erickson, E. F., Fich, M. 1997, ApJ, 489, 94
\bibitem{}
Scalo, J.M. 1986, Fundam. Cosmic Phys., 11, 1
\bibitem[]{}
Schaller, G., Schaerer, D., Meynet, G., \& Maeder, A. 1992, A\&AS, 96, 269
\bibitem{}
Simpson, J. P., Colgan, S. W. J., Rubin, R. H., Erickson, E. F., Haas, M. R. 1995, ApJ, 444, 721
\bibitem[]{}
Smith, V.V., Plez, B., Lambert, D.L., \& Lubowich, D.A. 1995, ApJ, 441, 735
\bibitem{}
Sneden, C. \& Primas, F. 2001, New Astron. Rev., 45, 513
\bibitem{}
Sofia, U. J., Meyer, D. M. 2001, ApJ, 554, L221
\bibitem[]{}
Thielemann, F. -K., Nomoto, K., \& Hashimoto, M. 1996, ApJ, 460, 408 (TNH)
\bibitem[]{}
Thielemann, F. -K., Nomoto, K. \& Hashimoto, M. 1993, in Origin and Evolution
of the Elements, ed. N. Prantzos et al., Cambridge University Press, p. 297
\bibitem{}
Thomas, D., Greggio, L., Bender, R. 1998, MNRAS, 296, 119
\bibitem{}
Thuan, T. X., Lecavalier des Etangs, A., Izotov, Y. I. 2002, ApJ, 565, 875
\bibitem[]{}
Thurston, T.R., Edmunds, M.G., \& Henry, R.B.C. 1996, MNRAS, 283, 990
\bibitem[]{}
Tsamis, Y. G., Barlow, M. J., Liu, X. -W., Danziger, I. J., Storey, P. J. 2002, MNRAS, in press
\bibitem[]{}
van den Hoek, L.B., \& Groenewegen, M.A.T. 1997, A\&AS, 123, 305 (vdHG)
\bibitem{}
van der Kruit, P. C. 1986, A\&A, 157, 230
\bibitem[]{}
van Zee, L., Salzer, J.J., Haynes, M.P., O'Donoghue, A.A., \& Balonek, T.J. 
   1998a, AJ, 116, 2805
\bibitem{}
van Zee, L., Salzer, J.J., Haynes, M.P. 1998b, ApJ, 497, L1
\bibitem{}
van Zee, L., Haynes, M. P., Salzer, J. J. 1997, AJ, 114, 2497
\bibitem{}
Vilchez, J. M., Esteban, C. 1996, MNRAS, 280, 720
\bibitem{}
Wheeler, J. C., Sneden, C. Truran, J. W. 1989, ARA\&A, 27, 279
\bibitem{}
Wielen, R., Fuchs, B., Dettbarn, C. 1996, A\&A, 314, 438
\bibitem[]{} 
Woosley, S. E., Weaver, T. A. 1995, ApJS, 101, 181 (WW)
\bibitem{}
Woosley, S. E., Axelrod, T. S., Weaver, T. A. 1984, in ``Stellar Nucleosynthesis'', eds. C. Chiosi and A. Renzini, Reidel (Dordrecht), p. 263
\bibitem{}
Zaritsky, D., Kennicutt, R. C., Huchra, J. P. 1994, ApJ, 420, 87
\end{thebibliography}
\end{document}